\documentclass[12pt]{article}
\usepackage{lscape}
\usepackage{latexsym}
\usepackage{amsmath,amsbsy,amssymb,bm,mathrsfs}
\usepackage{colordvi,multicol}
\usepackage{graphicx}
\usepackage{epstopdf}
\usepackage{epsfig}
\usepackage{color,xcolor}
\usepackage{float}
\usepackage{subfigure}
\usepackage{path}
\usepackage{url}
\usepackage{verbatim}
\usepackage{rotating}
\usepackage{natbib}
\usepackage{enumerate}
\usepackage{booktabs}
\usepackage{multirow,graphicx,lscape,float,threeparttable}
\usepackage{setspace}
\usepackage[ruled]{algorithm2e}
\usepackage{algorithmic}

\usepackage[bookmarks=true,
colorlinks,%
linkcolor=blue,%
citecolor=blue,%
plainpages=false,%
pdfstartview=FitH
]{hyperref}
\bibpunct{(}{)}{;}{a}{}{,}

\usepackage {xr}
\DeclareMathOperator*{\argmin}{arg\,min}

\newcommand{\uzero}    {\mbox{\boldmath$0$}}
\newcommand{\uone}     {\mbox{\boldmath$1$}}

\newcommand{\uB}       {\mbox{\boldmath$B$}}

\newcommand{\uh}       {\mbox{\boldmath$h$}}
\newcommand{\uI}       {\mbox{\boldmath$I$}}

\newcommand{\ut}       {\mbox{\boldmath$t$}}

\newcommand{\uu}       {\mbox{\boldmath$u$}}
\newcommand{\uV}       {\mbox{\boldmath$V$}}
\newcommand{\uv}       {\mbox{\boldmath$v$}}

\newcommand{\uw}       {\mbox{\boldmath$w$}}
\newcommand{\uX}       {\mbox{\boldmath$X$}}
\newcommand{\ux}       {\mbox{\boldmath$x$}}

\newcommand{\uy}       {\mbox{\boldmath$y$}}

\newcommand{\uz}       {\mbox{\boldmath$z$}}

\newcommand{\ubeta}             {\mbox{\boldmath$\beta$}}
\newcommand{\ugamma}            {\mbox{\boldmath$\gamma$}}

\newcommand{\uzeta}             {\mbox{\boldmath$\zeta$}}
\newcommand{\ueta}              {\mbox{\boldmath$\eta$}}

\newcommand{\uiota}             {\mbox{\boldmath$\uiota$}}

\newcommand{\ueps}              {\mbox{\boldmath$\eps$}}

\newcommand{\wcon}{\stackrel{{\cal L}}{\longrightarrow}}

\newcommand{\eps}{\varepsilon}

\def\calS{\mathcal{S}}
\DeclareMathOperator{\E}{\mathbb{E}}

\DeclareMathOperator{\tr}{\mbox{tr}}

\newcommand{\qed}{\hfill\hbox{\vrule height1.5ex width.5em}}

\def\l{\left}
\def\r{\right}

\def\trans{^{\rm T}}

\def\hubeta{\widehat{\boldsymbol\beta}}
\def\hugamma{\widehat{\boldsymbol\gamma}}

\newtheorem{theorem}{Theorem}

\newcommand{\blind}{1}

\date{}
\begin{document}

	\def\spacingset#1{\renewcommand{\baselinestretch}%
		{#1}\small\normalsize} \spacingset{1}

	
	\if1\blind
	{
		\title{\bf Testing relevant difference in high-dimensional linear regression with applications to detect transferability}
		\author{
			Xu Liu\\
            Shanghai University of Finance and Economics
		}
		\maketitle
	} \fi
	
	\if0\blind
	{
		\bigskip
		\bigskip
		\bigskip
		\begin{center}
			{\LARGE\bf Testing relevant difference in high-dimensional linear regression with applications to detect transferability}
		\end{center}
		\medskip
	} \fi
	
\bigskip
\begin{abstract}
Most of researchers on testing a significance of coefficient $\ubeta$ in high-dimensional linear regression models consider the classical hypothesis testing problem $H_0^{c}: \ubeta=\uzero \mbox{ versus } H_1^{c}: \ubeta \neq \uzero$. We take a different perspective and study the testing problem with the null hypothesis of no relevant difference between $\ubeta$ and $\uzero$, that is, $H_0: \|\ubeta\|\leq \delta_0 \mbox{ versus } H_1: \|\ubeta\|> \delta_0$, where $\delta_0$ is a prespecified small constant. This testing problem is motivated by the urgent requirement to detect the transferability of source data in the transfer learning framework. We propose a novel test procedure incorporating the estimation of the largest eigenvalue of a high-dimensional covariance matrix with the assistance of the random matrix theory. In the more challenging setting in the presence of high-dimensional nuisance parameters, we establish the asymptotic normality for the proposed test statistics under both the null and alternative hypotheses. By applying the proposed test approaches to detect the transferability of source data, the unified transfer learning models simultaneously achieve lower estimation and prediction errors with comparison to existing methods. We study the finite-sample properties of the new test by means of simulation studies and illustrate its performance by analyzing the GTEx data.
\end{abstract}

\noindent%
\vfill

\newpage
\spacingset{1.9} 

\section{Introduction}
It is a fundamental problem in statistical learning of high-dimensional linear regression models to test the significance of a high-dimensional vector. Specifically, given control factors that are known to be correlated with the response, it is important to accurately and efficiently test the significance of a high-dimensional subvector of the model coefficients corresponding to the rest covariates. A lot of literature focus on the high-dimensional test to detect whether the entire coefficients are zero or not, in which the crucial idea comes originally from the two-sample test. \cite{ChenQin2010} considered the two-sample mean test, and replaced the covariance matrix with the identity matrix to avoid issues in the case of ``large $p$ small $n$'' as considered by \cite{BaiSaranadasa1996}. Other works, such as \cite{ZhongChen2011,CuiGuoZhong2018,ZhangYaoShao2018}, applied this technique in the linear regression
coefficient test. \cite{Chen2024} studied the hypothesis test of high-dimensional variables on quantile regression with low-dimensional control factors, in which the authors proposed a novel test statistic by constructing a U-statistic based on score function. However, all these methods are not applicable in the presence of high-dimensional control factors. \cite{Chen2023} proposed a U-test statistic for generalized linear models in the presence of high-dimensional control factors, which extends previous work by \cite{GuoChen2016} in which they handle low-dimensional control factors. \cite{Yang2023} proposed a score-based test for ultrahigh-dimensional linear regression model in the presence of high-dimensional control factors. More references can be found in \cite{Ning2017,LiLi2022}. \cite{Liu2024} applied the testing method proposed by \cite{Chen2023} to detect transferability of the source data in transfer learning frameworks.

This work is motivated by the following fact: borrowing information from useful source data improves the efficiency of fitting on target data in the transfer learning framework if the difference between parameters in the models fitting target and source data is quite small. According to \cite{Tian2023, Li2022FDR,Li2022Minimax, LiShen2023, LiZhang2024}, when $\delta=\|\ubeta_k-\ubeta_0\|_2$ is small enough, such as $\delta \ll \sqrt{sn_0^{-1}\log p}$, the upper bounds of their method are better than the classical Lasso bound $\mathcal{O}_p(\sqrt{sn_0^{-1}\log p})$ only on target data, where $\ubeta_0$ and $\ubeta_k$ are $p$-dimensional coefficients of the target and the $k$th source models, respectively. In other words, it is helpful to fit the target data by borrowing the source data if $\ubeta_k$ is in the neighborhood of $\ubeta_0$ with a small radius $\delta$, instead of the strict requirement $\ubeta_k=\ubeta_0$.
This fact implies that it is necessary to relax the hypothesis testing problem
\begin{align}\label{test_classic}
	H_0^{c}: \ubeta=\uzero \quad \mbox{versus}\quad H_1^{c}: \ubeta \neq \uzero,
\end{align}
where $\ubeta=\ubeta_k-\ubeta_0$.
Thus, instead of testing problem (\ref{test_classic}), we consider
testing the relevant difference versus no small relevant difference, that is,
\begin{align}\label{test}
	H_0: \|\ubeta\|\leq \delta_0 \quad \mbox{versus}\quad H_1: \|\ubeta\|> \delta_0,
\end{align}
where $\delta_0$ is a prespecified constant that represents the maximal difference of coefficients in regression models such that the source data is helpful for fitting target data.

Hypothesis test of the relevance in (\ref{test}) is also called precise hypothesis (\cite{Berger1987}) or approximate hypothesis (\cite{Hodges1954, Rosenblatt1962a, Rosenblatt1962b, BergerSellke1987, Berger1996}) in the literature, which has nowadays been widely recognized in various fields of statistical inference including medical, pharmaceutical, chemistry and environmental statistics (see \cite{ChowLiu2008}, \cite{McBride1999}, \cite{Callegaro2019}, and \cite{Lehmann2022}). A relevant change in the change point analysis of the time series  was studied by \cite{Dette2016}. \cite{Dette2020Testing} studied testing relevant hypotheses for functional time series.
\cite{Dette2020Functional, Bucher2021} and \cite{Dette2024} investigated the hypothesis test for relevant differences for functional data. \cite{Bastian2024} developed a test procedure for testing the mutual independence of the components of a high-dimensional vector with the null hypothesis that all pairwise associations do not exceed a certain threshold in absolute value.
To our best knowledge, however, the problem of testing for relevant difference in high-dimensional regression models has not been discussed in the literature so far. As pointed out by \cite{Berkson1938}, the hypothesis testing problem (\ref{test}) avoids the consistency problem, that is, any consistent test will detect any arbitrary small change in the parameters if the sample size is sufficiently large.

The considered testing problem (\ref{test}) leads to an immediate application to detect transferability of source data in the transfer learning. Transfer learning has been widely utilized in medical and biological studies, such as predictions of protein localization (\cite{Mei2011}), biological imaging diagnosis (\cite{Shin2016}), drug sensitivity prediction (\cite{Turki2017}) and integrative analysis of ``multi-omics" data (\cite{SunHu2016}, \cite{Hu2019}, and \cite{Wang2019}). More details can be found in  the recent survey papers (\cite{PanYang2009,Zhuang2020}). Recently, many authors focused on development of methodological and theoretical works for transfer learning. \cite{Tong2021} established the transferability measure on the discrete and continuous data by the $\chi^2$-distance between the source and target data. \cite{Tripuraneni2020} provided a two-stage empirical risk minimization procedure for transfer learning and derived generalization bounds with general losses, tasks, and features. \cite{Li2022Minimax} focused on prediction, estimation and minimax optimality of transfer learning for high-dimensional linear regressions. \cite{CaiWei2021} established the minimax rate of convergence and constructed a rate optimal two-sample weighted K-NN classifier by transfer learning. \cite{Liu2024} studied the unified transfer learning models for high-dimensional linear regression. More literature on transfer learning in high-dimensional models can be referred to \cite{Bastani2021,Tian2023,Li2022FDR,LiShen2023,LiZhang2024}.
These papers either assume that the transferable source data is known or detect the transferability of source data by applying test procedure to hypothesis testing problem (\ref{test_classic}) instead of (\ref{test}). This may be too strict to detect transferable source data in transfer learning according to their theoretical properties.

In this paper, we first investigate the test procedure for relevant difference (\ref{test}) in high-dimensional linear regression models, where we use random matrix theory to estimate the largest eigenvalue of a high-dimensional covariance that is a crucial ingredient in our procedure. It is well-known that estimating the largest eigenvalue of high-dimensional covariance is a big challenge.
We provide two algorithms to solve the spectrum estimation based on the Mar\v{c}enko-Pastur equation (\cite{Karoui2008} and \cite{Kong2017}) with moment constraints.
In more challenging settings where the nuisance coefficients in high-dimensional linear regression models are also high-dimensional, we derive asymptotic normality of the proposed test statistics. 
The theoretical and numerical results demonstrate that the proposed test statistic is powerful to test the relevant difference of coefficients.
By applying the proposed test method to the transfer learning, the unified transfer learning models simultaneously achieve lower estimation and prediction errors, which demonstrates the proposed testing methods efficiently detect the transferable source data. Two efficient algorithms are designed to implement the proposed test methods and transfer learning.

The rest of the paper is organized as follows. In Section \ref{sec:test}, we investigate the test procedure for testing relevant difference in high-dimensional linear regression models. We provide estimation procedures for the largest eigenvalue, and build the asymptotic normality of proposed test statistic under regular conditions. Section \ref{sec:test_trans} provides the testing procedure for detecting the transferability of source data. In Section \ref{sec:sim_test}, we conduct simulation studies to evaluate the finite-sample performance of the proposed test statistics and transfer learning procedures. In Section~\ref{sec:case_studies}, we apply the proposed method to the GTEx data. Finally, in Section~\ref{sec:conclusion}, we conclude with remarks and further extensions. Conditions, proofs and additional simulation studies are included in the Supplementary Material.
An R package named ``hdtrd'' is available at \url{https://github.com/xliusufe/hdtrd}.

Let $\mathbb{R}^p$ denote the set of real numbers with dimension $p$, $\mathbb{C}$ denote the set of complex numbers, $\mbox{RE}(z)$ and $\mbox{IM}(z)$ denote the real and imaginary part of $z\in\mathbb{C}$, respectively. Denote by $\uI_p\in\mathbb{R}^{p\times p}$ the identity matrix, and by $\uone(\cdot)$ the indicator function. For a vector $\uv=(v_1,\cdots,v_d)\trans\in \mathbb{R}^{d}$, denote by $\|\uv\|$ the Euclidean norm of $\uv$, by $\|\uv\|_1=\sum_{j=1}^d|v_j|$ the $L_1$ norm, and by $\|\uv\|_0=\sum_{j=1}^d\uone(v_j\neq0)$ the $L_0$ norm. For a square matrix $A\in\mathbb{R}^{d\times d}$, denote by $\lambda_{\max}(A)$ and $\lambda_{\min}(A)$ the maximum and minimum eigenvalue of $A$, and by $\tr(A)$ the trace of $A$.

\section{Test of relevant difference}\label{sec:test}

Let $y$ be the response variable, $\ux$ be $p_1$-dimensional covariates, and $\uz$ be $p_2$-dimensional control factors.
Given independent and identically distributed (i.i.d.) samples $(y_i, \ux_i, \uz_i)$, $i = 1, \ldots, n$, we consider the linear regression model
\begin{align}\label{reg.1}
	y_i = \ux_{i}\trans\ubeta+\uz_i\trans\ugamma+\eps_i,
\end{align}
where $\ubeta\in \mathbb{R}^{p_1}$ is the parameter of interest, $\ugamma\in \mathbb{R}^{p_2}$ is the nuisance parameter, $p=p_1+p_2$, and the error term $\eps_i$ is independent of $(\ux_i\trans,\uz_i\trans)\trans$ satisfying
$\E\eps_i = 0$ and $\E\eps_i^2=\sigma^2<\infty$. Herein, we allow that both $p_1$ and $p_2$ are greater than sample size $n$ and diverge as $n$ goes to infinity.

Consider following hypothesis testing problem,
\begin{align}\label{test0}
	H_0: \|\ubeta-\ubeta_0\|\leq \delta_0 \quad \mbox{versus}\quad H_1: \|\ubeta-\ubeta_0\|> \delta_0,
\end{align}
where $\delta_0$ is a specified constant.
According to \cite{Tian2023}, \cite{LiShen2023} and \cite{LiZhang2024}, when
$\delta_0 \ll \sqrt{sn_0^{-1}\log p}$,
the upper bounds of their method $\mathcal{A}_{\delta_0}$-Trans are better than the classical Lasso bound $\mathcal{O}_p(\sqrt{sn_0^{-1}\log p})$ on target data only.
Without loss of generality, we can set $\ubeta_0=\uzero$, that is, we can consider the testing problem (\ref{test}). 
It is seen that when $\delta_0=0$, the testing problem (\ref{test}) reduces to the classic testing problem (\ref{test_classic}),
which has been well studied in the literature, 
such as \cite{GuoChen2016}, \cite{CuiGuoZhong2018}, \cite{Chen2023} and \cite{Yang2023}.

\subsection{Test statistic when $\delta_0=0$}\label{sec:test_old}

In this section, we review the classic testing problem (\ref{test_classic}).
When $\uz$'s dimension $p_2$ diverges but is low, \cite{GuoChen2016} proposed a test statistic
\begin{align}\label{Tnp}
	T_{n,p}^{0}
	=\frac{1}{n(n-1)}	\sum_{i\neq j}\ux_i\trans\ux_j(y_i-\uz_i\trans\hugamma)(y_j-\uz_j\trans\hugamma),
\end{align}
where $\hugamma$ is a consistent estimator of $\ugamma$ under the null.
\cite{Chen2023} extended test statistic (\ref{Tnp}) to high-dimensional $\uz$, where $p_2>n$ diverges. They replaced $\hugamma$ by
minimizing the penalized loss function with Lasso penalty \citep{Tibshirani1996}
\begin{align}\label{obj_gamma}
	\hugamma=\argmin_{\ugamma\in R^{p_2}}\frac{1}{2n}\sum_{i=1}^n\l(y_i - \uz_i\trans\ugamma\r)^2+\lambda\|\ugamma\|_1.
\end{align}

For the testing problem (\ref{test_classic}), \cite{Chen2023} and \cite{Yang2023} showed that the test statistic $T_{n,p}^{0}$ is asymptotically normally distributed under some conditions. The following Theorem \ref{th:T0} needs some notation. Denote $\Sigma_{xz}=\E\ux\uz\trans\in\mathbb{R}^{p_1\times p_2}$ and by $\Sigma_z\in\mathbb{R}^{p_2\times p_2}$ the covariance of $\uz$. Define a family of coefficient $\ubeta$
\begin{align*}
	\mathfrak{F}=\left\{\ubeta\in\mathbb{R}^{p_1}:\ubeta\trans\Sigma_{\eta}\ubeta=o(1),\ubeta\trans\Sigma_{\eta}^2\ubeta=o\left(\frac{\tr(\Sigma_{x}^2)}{\varsigma_n^2}\right), \ubeta\trans\Sigma_{\eta}\Sigma_{x}\Sigma_{\eta}\ubeta=o\left(\frac{\tr(\Sigma_{x}^2)}{n}\right)\right\},
\end{align*}
where $\Sigma_{\eta}=\Sigma_x-\Sigma_{xz}\Sigma_z^{-1}\Sigma_{xz}\trans$ is the covariance of $\ueta=\ux-\Sigma_{xz}\Sigma_z^{-1}\uz$, $\varsigma_n=ns_{\phi}\log p_2$ with $s_{\phi}=\|\ugamma_{\phi}\|_0$, and $\ugamma_{\phi}$ is defined in Lemma A1 of the Supplementary Material.
\begin{theorem}\label{th:T0}
	If Conditions {\rm(\textbf{A.1})-(\textbf{A.5})} in the Supplementary Material hold, it follows that for any given $\ubeta\in\mathfrak{F}$,
	\begin{align*}
		\frac{n(T_{n,p}^{0}-\ubeta\trans\Sigma_{\eta}^2\ubeta)}{\sigma^2\sqrt{2\tr(\Sigma_x^2)}}\wcon \mathcal{N}(0,1),
	\end{align*}
	where $\wcon$ stands for convergence in distribution.
\end{theorem}
Theorem \ref{th:T0} states that one can reject $H_0^c$ at a significant level $\alpha$ if $T_{n,p}^{0}>\frac{z_{\alpha}}{n}\sqrt{2\widehat{\sigma^4\tr(\Sigma_x^2)}}$,
where $z_{\alpha}$ is the upper $\alpha$-quantile of standard normal distribution, and $\widehat{\sigma^4\tr(\Sigma_x^2)} = \frac{1}{n(n-1)}\sum_{i\neq j}^n (\ux_i^T\ux_j)^2(y_i-\uz_i\trans\hugamma)^2(y_j-\uz_j\trans\hugamma)^2$ proposed by \cite{GuoChen2016} is a consistent estimator of $\sigma^4\tr(\Sigma_x^2)$.
Here, $\hugamma$ is a consistent estimator of $\ugamma$ given in (\ref{obj_gamma}).

From Lemma B1 in the Supplementary Material, the asymptotic normality in Theorem \ref{th:T0} requires the weak correlation between $\uz$ and $\ux$, which is imposed in Condition (A.5)  $\varrho_n\lambda_{\max}^{1/2}(\Sigma_x)\{s_{\phi}a_n\log n\log p_2\}^{1/2}=o(\tr^{1/2}(\Sigma_x^2))$, where 
$\varrho_n=\max_{k\in\{1,\cdots,p_2\}}\|\E\ux z_k\|$ with $z_k$ being the $k$th element of $\uz$. $\varrho_n$ could be $O(p)$ if $z_k$ correlates highly with $\ux$, which results in $\lambda_{\max}^{1/2}(\Sigma_x)\{s_{\phi}a_n\log n\log p_2\}^{1/2}>n^{1/2}p_2^{-1}\tr^{1/2}(\Sigma_x^2)$ and may break Condition (A.5).

It is necessary to reduce the correlation between $\uz$ and $\ux$, which can be realized by the regularized projection score method \citep{Cheng2022}. Specifically, we consider the projection score for $\ugamma$ as $S(\ugamma) = -\uz(y-\ux\trans\ubeta-\uz\trans\ugamma)$.
The projection score is defined as the residual of the projection of the score function $S(\ubeta)=-\ux(y-\ux\trans\ubeta-\uz\trans\ugamma)$ onto the closure of the linear span of the score function $S(\ugamma)$ in the Hilbert space $L_2(P)$, where $P$ is the distribution of $(y,\ux,\uz)$. Thus, we need to find a matrix $H\in\mathbb{R}^{p_1\times p_2}$ that minimizes $\E\|S(\ubeta)-H S(\ugamma)\|^2=\sigma^2\sum_{j=1}^{p_1}\E(x_j-\uz\trans \uh_j)^2$,
where $\uh_j\in\mathbb{R}^{p_2}$ is the $j$th column of $H\trans$. In particular, $H$ satisfies the normal equation $\E(\ux-H\uz)\uz\trans=\uzero$, which yields $H=\Sigma_{xz}\Sigma_{z}^{-1}$ if $\Sigma_z$ is invertible.

However, the sample version of $\Sigma_z$, which is estimated traditionally by $n^{-1}\sum_{i=1}^{n}\uz_i\uz_i\trans$, is not invertible when $p_2>n$. Therefore, it is needed to regularize the projection calculation, in which we apply the standard Lasso \citep{Cheng2022}
\begin{align}\label{hatH}
	\widehat{H} = \argmin_{H\in\mathbb{R}^{p_1\times p_2}}\frac{1}{2n}\sum_{i=1}^{n}\|\ux_i-H\uz_i\|^2+\lambda\sum_{j=1}^{p_1}\|\uh_j\|_1.
\end{align}
Based on the residual of the projection score $\hat{\ueta}_i=\ux_i-\widehat{H}\uz_i$ for $i=1,\cdots,n$, we construct a test statistic $T_{n,p}^{0,Proj}$ as
\begin{align}\label{Tnp_project}
	T_{n,p}^{0,Proj}
	=\frac{1}{n(n-1)}
	\sum_{i\neq j}\hat{\ueta}_i\trans\hat{\ueta}_j(y_i-\uz_i\trans\hugamma)(y_j-\uz_j\trans\hugamma).
\end{align}
Define a family of coefficients as $\mathfrak{F}^{Proj}=\left\{\ubeta\in\mathfrak{F}:\ubeta\trans\Sigma_{\eta}^2\ubeta=o\left(\varsigma_n^{-1}\tr(\Sigma_{\eta}^2)\right)\right\}$.
We show that the test statistic $T_{n,p}^{0,Proj}$ is asymptotically normally distributed in the following Theorem \ref{th:T0_proj}.
\begin{theorem}\label{th:T0_proj}
	If Conditions {\rm(\textbf{A.1})-(\textbf{A.7})} in the Supplementary Material hold,
	it follows that for any given $\ubeta\in\mathfrak{F}^{Proj}$,
	\begin{align*}
		\frac{T_{n,p}^{0,Proj}-\ubeta\trans\Sigma_{\eta}^2\ubeta}{n^{-1}\sigma^2\sqrt{2\tr(\Sigma_{\eta}^2)}}\wcon \mathcal{N}(0,1).
	\end{align*}
\end{theorem}
\cite{Yang2023} has provided the same test statistic $T_{n,p}^{0,Proj}$ as well as its asymptotic normality with the known projection $H$ or $\widehat{H}$ estimated based on another data independent of $\uV_n=\{(y_i,\ux_i,\uz_i)\}_{i=1}^n$. In this paper,
with the assistance of U-process \citep{Chen2020}, we establish the asymptotic normality of $T_{n,p}^{0,Proj}$ based on $\widehat{H}$ estimated from $\uV_n$.

To compare theoretically asymptotic power of test statistics between $T_{n,p}^{0,Proj}$ and $T_{n,p}^{0}$,
we define asymptotic relative efficiency (ARE) as $\mbox{ARE}(T_{n,p}^{0,Proj}, T_{n,p}^{0})
=\frac{\tr^{1/2}(\Sigma_x^2)}{\tr^{1/2}(\Sigma_{\eta}^2)}$.
Since $\Sigma_{xz}\Sigma_z^{-1}\Sigma_{xz}\trans$ is positive-definite, by the definition $\Sigma_{\eta}=\Sigma_x-\Sigma_{xz}\Sigma_z^{-1}\Sigma_{xz}\trans$, it is seen that $\tr(\Sigma_{\eta}^2)\leq\tr(\Sigma_x^2)$, which results in $\mbox{ARE}(T_{n,p}^{0,Proj}, T_{n,p}^{0})\geq1$. Thus, projection test statistic $T_{n,p}^{0,Proj}$ achieves a greater asymptotic power than the competitor $T_{n,p}^{0}$.

From Theorem \ref{th:T0_proj}, one can reject $H_0^c$ in testing problem (\ref{test_classic}) at a significant level $\alpha$ if
\begin{align}\label{reject0_proj}
	T_{n,p}^{0,Proj}>\frac{z_{\alpha}}{n}\sqrt{2\widehat{\sigma^4\tr(\Sigma_{\eta}^2)}},
\end{align}
where $\widehat{\sigma^4\tr(\Sigma_{\eta}^2)} = \frac{1}{n(n-1)}\sum_{i\neq j}^n (\ueta_i\trans\ueta_j)^2(y_i-\uz_i\trans\hugamma)^2(y_j-\uz_j\trans\hugamma)^2$.

On the other hand, Theorem \ref{th:T0_proj} implies that the rejection region (\ref{reject0_proj}) can not control the type-I error for testing problem (\ref{test}). The power function for testing problem (\ref{test_classic}) is
$\Omega_n^{0}(\ubeta)=\Phi\left(-z_{\alpha}+n\ubeta\trans\Sigma_{\eta}^2\ubeta(2\sigma^4\tr(\Sigma_{\eta}^2))^{-1/2}\right)\{1+o(1)\}$, where $\Phi(\cdot)$ is the cumulative distribution function of $\mathcal{N}(0,1)$.
The inflation comes from the signal-to-noise ratio (SNR) $d_n^{0}(\ubeta)=n\ubeta\trans\Sigma_{\eta}^2\ubeta(2\sigma^4\tr(\Sigma_{\eta}^2))^{-1/2}$.
To control the type-I error, it is natural to construct an efficient test statistic by subtracting $\sup_{\|\ubeta\|\leq\delta_0}d_n^{0}(\ubeta)$ from $T_{n,p}^{0,Proj}$. This observation motivates us to construct a new test statistic for testing problem (\ref{test}), which is investigated in Section \ref{sec:eigmax}.

\subsection{Test statistic when $\delta_0>0$}\label{sec:eigmax}
Now we are ready to consider the testing problem (\ref{test}). Following arguments in previous section, we have  $\sup_{\|\ubeta\|\leq\delta_0}d_n^{0}(\ubeta)=\delta_0^2\lambda_{\max}^2(\Sigma_{\eta})\left\{\sigma^2\sqrt{2\tr(\Sigma_{\eta}^2)}\right\}^{-1}$.
According to Theorem \ref{th:T0_proj}, we can observe that
under $H_0$ in testing problem (\ref{test})
\begin{align*}
	\lim_{n\rightarrow\infty}P\left(T_{n,p}^{0,Proj}-\delta_0^2\lambda_{\max}^2(\Sigma_{\eta})>\sigma^2\sqrt{2\tr(\Sigma_{\eta}^2)}z_{\alpha}/n\right)
	\leq \alpha,
\end{align*}
and under $H_1$
{\small
	\begin{align*}
		\lim_{n\rightarrow\infty}&P\left(T_{n,p}^{0,Proj}-\delta_0^2\lambda_{\max}^2(\Sigma_{\eta})>\sigma^2\sqrt{2\tr(\Sigma_{\eta}^2)}z_{\alpha}/n\right)
		=\begin{cases}
			\alpha,& \hbox{if } ~n\frac{\ubeta\trans\Sigma_{\eta}^2\ubeta-\delta_0^2\lambda_{\max}^2(\Sigma_{\eta})}{\sigma^2\sqrt{2\tr(\Sigma_{\eta}^2)}}=o(1), \\
			1, & \hbox{if } ~n\frac{\ubeta\trans\Sigma_{\eta}^2\ubeta-\delta_0^2\lambda_{\max}^2(\Sigma_{\eta})}{\sigma^2\sqrt{2\tr(\Sigma_{\eta}^2)}}\rightarrow+\infty.	\end{cases}
	\end{align*}
}
Finally, we reject $H_0$ in (\ref{test}) at a significant level $\alpha$ if
$T_{n,p}^{0,Proj}>\delta_0^2\widehat{\lambda_{\max}^2(\Sigma_{\eta})}+\frac{z_{\alpha}}{n}\sqrt{2\widehat{\sigma^4\tr(\Sigma_{\eta}^2)}}$,
where $\widehat{\sigma^4\tr(\Sigma_{\eta}^2)}$ is defined as in Section \ref{sec:test_old}, and $\widehat{\lambda_{\max}^2(\Sigma_{\eta})}$ is a consistent estimator of $\lambda_{\max}^2(\Sigma_{\eta})$. This observation encourages
us to propose a new test statistic for testing problem (\ref{test})
\begin{align}\label{statistic}
	T_{n,p} = T_{n,p}^{0,Proj}-\delta_0^2\widehat{\lambda_{\max}^2(\Sigma_{\eta})}.
\end{align}

In Appendix F of the Supplementary Material, we provide an estimator $\widehat{\lambda_{\max}(\Sigma_{\eta})}=\alpha_n(1+\tau_n)^{-2}\lambda_{\max}(S_n)$ for the largest eigenvalue of $\Sigma_{\eta}$. It is convenient to implement, but is not consistent, and consequently brings bias in practice. In this section, we provide two consistent estimators for the largest eigenvalue of $\Sigma_{\eta}$.

\cite{Karoui2008} proposed an algorithm to solve the spectrum estimation problem by the Mar\v{c}enko-Pastur equation
\begin{align}\label{mplaw}
	z=-\frac{1}{\underline{m}_{G}(z)}+\tau^2\int\frac{t}{1+t\underline{m}_{G}(z)}dG(t), ~\forall z\in\mathbb{C}^+,
\end{align}
where $\tau=\lim p/n$, $\mathbb{C}^+=\{z\in\mathbb{C}:\mbox{IM}(z)>0\}$, and $\underline{m}_{G}(z)$ is the Stieltjes transform for the spectral distribution $G(\cdot)$. $\underline{m}_{G}(z)$ can be approximated  by the Stieltjes transform for the empirical spectral distribution $G_n(\cdot)$, that is,
\begin{align}\label{stieltjes}
	\underline{m}_{G}(z)\simeq \underline{m}_{G_n}(z)= -\frac{1-p_1/n}{z}+\frac{1}{n}\tr\l((S_n-z\uI_{p_1})^{-1}\r).
\end{align}
By applying discretization strategy, we approximate the integral in Mar\v{c}enko-Pastur equation (\ref{mplaw}) by
\begin{align}\label{mplaw_dis}
	\int\frac{t}{1+t\underline{m}_{G}(z)}dG(t)
	\simeq\sum_{j=1}^{J_t}\omega_j\frac{t_j}{1+t_j\underline{m}_{G}(z)},
\end{align}
where $J_t$ is a positive integer, and $\{\omega_j,j=1,\cdots,J_t\}$ are weights with the constraints $\sum_{j=1}^{J_t}\omega_j=1$ and $\omega_j\geq0$.
Thus, following above arguments, Mar$\check{c}$enko-Pastur equation (\ref{mplaw}) turns into
\begin{align}\label{mplaw_approx}	z\simeq-\frac{1}{\underline{m}_{G_n}(z)}+\frac{p_1}{n}\sum_{j=1}^{J_t}\omega_j\frac{t_j}{1+t_j\underline{m}_{G_n}(z)}, \forall z\in\mathbb{C}^+,
\end{align}
where $\{t_j\in[a,b],j=1,\cdots,J_t\}$ with $t_1\leq\cdots\leq t_{J_t}$ is a set of discrete points, and $[a,b]\subset\mathbb{R}$ is a preset interval.
To approximate the largest eigenvalue of $\Sigma_{\eta}$, it suffices to find $t_{j_0}\in\{t_j,j=1,\cdots,J_t\}$ with $j_0=\min\{j:\sum_{\ell=1}^{j}\omega_{\ell}\geq p_1/(p_1+1), j=1,\cdots,J_t\}$. Thus, it is crucial to find weights $\{\omega_j,j=1,\cdots,J_t\}$.

Recall $t_j\in[a,b]$ with $t_1\leq\cdots\leq t_{J_t}$. In practice, we take $a=\lambda_{\min}(S_n)/\lambda_{\max}(S_n)$ and $b=\lambda_{\max}(S_n)$. To solve $\{\omega_j,j=1,\cdots,J_t\}$ from equation (\ref{mplaw_approx}), we set $z_k=u_k+\frac{1}{\sqrt{n}}i$, $k=1,\cdots,J_z$, where $i=\sqrt{-1}$ stands for the imaginary unit, $u_k\in\mathbb{R}$, and $J_z>J_t$ is a positive integer. In practice, we generate $u_k$ randomly from standard normal distribution $\mathcal{N}(0,1)$.
Invoking that $\mbox{RE}(z)$ and $\mbox{IM}(z)$ denote the real and imaginary parts of $z\in\mathbb{C}$, respectively. Denote $y_{R,k}=\mbox{RE}(z_k+1/\underline{m}_{G_n}(z_k))$, $y_{I,k}=\mbox{IM}(z_k+1/\underline{m}_{G_n}(z_k))$, and
\begin{align*}
	x_{R,k,j}=\mbox{RE}\l(\frac{p_1t_j/n}{1+t_j\underline{m}_{G_n}(z_k)}\r),\mbox{ and }
	x_{I,k,j}=\mbox{IM}\l(\frac{p_1t_j/n}{1+t_j\underline{m}_{G_n}(z_k)}\r).
\end{align*}
To find weights $\uw=(\omega_1,\cdots,\omega_{J_t})\trans$, we define a loss function of weights as follows,
\begin{align}\label{loss_omega}
	L_{\mbox{eig}}(\uw)=\frac{1}{J_t}\sum_{k=1}^{J_z}\left|y_{R,k}-\sum_{j=1}^{J_t}x_{R,k,j}\omega_j\right|+\frac{1}{J_z}\sum_{k=1}^{J_z}\left|y_{I,k}-\sum_{j=1}^{J_t}x_{I,k,j}\omega_j\right|
\end{align}
with the constraints $\sum_{j=1}^{J_t}\omega_j=1$ and $\omega_j\geq0$, $j=1,\cdots,J_t$.

\cite{Tian2015} proposed the unbiased and consistent estimators of the first four moments of population spectrum $G(t)$. Let $\zeta_k=\int t^kdG(t)$ and $\hat{\xi}_k=\int t^kdG_n(t)$ be the $k$th moment of population spectral distribution $G(t)$ and its empirical version $G_n(t)$, respectively. The unbiased and consistent estimators of the first four moments are constructed by $\hat{\zeta}_1 = \hat{\xi}_1$,
$\hat{\zeta}_2 = c_{2n}(\hat{\xi}_2-\tau_n^2\hat{\xi}_1)$,
$\hat{\zeta}_3 = c_{3n}(\hat{\xi}_3-3\tau_n^2\hat{\xi}_2\hat{\xi}_1+2\tau_n^4\hat{\xi}_1^3)$,
$\hat{\zeta}_4=c_{4n}(\hat{\xi}_4-4\tau_n^2\hat{\xi}_3\hat{\xi}_1-\tau_n^2\hat{\xi}_2^2\frac{2n^2+3n-6}{n^2+n+2}+\tau_n^4\hat{\xi}_1^2(2\hat{\xi}_2-\tau_n^2\hat{\xi}_1^2)\frac{5n^2+6n}{n^2+n+2})$,
where $\tau_n=\sqrt{p_1/n}$, $c_{2n}=n^2[(n-1)(n+2)]^{-1}$, $c_{3n}=c_{2n}n^2[(n-2)(n+4)]^{-1}$, and $c_{4n}=c_{3n}n(n^2+n+2)[(n-3)(n+1)(n+6)]^{-1}$. Thus, it is natural to consider following four moment constraints to improve the estimation efficiency $\hat{\zeta}_k=\sum_{j=1}^{J_t}\omega_jt_j^k$, $k=1,2,3,4$.
The optimization problem (\ref{loss_omega}) imposing above four moment constraints is convex, and is easily solved by classic algorithm in linear program. In fact, minimizing optimization problem (\ref{loss_omega}) is equivalent to following linear programming problem (LP),
\begin{align}\label{mplp}
	\begin{split}
		&\min_{\tilde{\uw}\in\mathbb{R}^{2J_z}}\uone\trans\tilde{\uw},\\
		&\mbox{subject to: } A\uw=\hat{\uzeta},\\
		&\left(
		\begin{array}{cc}
			-\ux_z & \uI_{2J_z} \\
			\ux_z & \uI_{2J_z} \\
		\end{array}
		\right)\left(
		\begin{array}{c}
			\uw \\
			\tilde{\uw} \\
		\end{array}
		\right)\geq\left(
		\begin{array}{c}
			-\uy_z \\
			\uy_z \\
		\end{array}
		\right),\\
		&~\mbox{and }\omega_j\geq0, j=1,\cdots,J_t,
	\end{split}
\end{align}	
where $\ux_z=(\ux_R\trans,\ux_I\trans)\trans$ 
with $\ux_R=(x_{R,k,j})_{k,j}\in\mathbb{R}^{J_z\times J_t}$ and $\ux_I=(x_{I,k,j})_{k,j}\in\mathbb{R}^{J_z\times J_t}$, and $\uy_z=(\uy_R\trans,\uy_I\trans)\trans$ with $\uy_R=(y_{R,1},\cdots,y_{R,J_z})\trans\in\mathbb{R}^{J_z}$, $\uy_I=(y_{I,1},\cdots,y_{I,J_z})\trans\in\mathbb{R}^{J_z}$, $A=(\uone,\ut,\ut^2,\ut^3,\ut^4)\trans\in\mathbb{R}^{5\times J_t}$ with $\ut=(t_1,\cdots,t_{J_t})\trans$, and $\hat{\uzeta}=(1,\hat{\zeta}_1,\hat{\zeta}_2,\hat{\zeta}_3,\hat{\zeta}_4)\trans\in\mathbb{R}^{5}$.
We employ the optimization package {\it limSolve} within R \citep{Meersche2009}. The estimator of the largest eigenvalue is given by
\begin{align}\label{eigenmax_hat}
	\widehat{\lambda_{\max}(\Sigma_{\eta})} = \min\left\{t_j: \sum_{\ell=1}^{j}\hat{\omega}_{\ell}\geq \frac{p_1}{p_1+1}, j=1,\cdots,J_t\right\},
\end{align}
which is the $p_1$th $(p_1+1)$st-quantile of distribution corresponding to $\hat{\uw}$, where $\hat{\uw}=(\hat{\omega}_1,\cdots,\hat{\omega}_{J_t})\trans$ is the estimator of $\uw$ by minimizing LP (\ref{mplp}).

\cite{Kong2017} introduced another way to estimate the largest eigenvalue of $\Sigma_{\eta}$. They proposed a consistent estimator of the $k$th moment of population spectral distribution, and then the consistent estimator of eigenvalue was provided based on the estimated moments. Compared with the method by \cite{Karoui2008}, a computationally cheaper algorithm was provided by \cite{Kong2017} to estimate the largest eigenvalue. 
\cite{Kong2017} proposed the $k$th moment of population distribution by
$\tilde{\zeta}_k = n^kp_1^{-1}\binom{n}{k}^{-1}\tr(\tilde{G}^{k-1}\tilde{S}_n)$,
where $\tilde{S}_n=n^{-1}\ueta\ueta\trans\in\mathbb{R}^{n\times n}$, and $\tilde{G}$ is the matrix $\tilde{S}_n$ with the diagonal and lower triangular entries set to zero. Then, they estimated the weight $\uw$ by solving following LP,
\begin{align}\label{mpmom}
	\begin{split}
		&\min_{\tilde{\uw}\in\mathbb{R}^{M}}\uone\trans\tilde{\uw},\\
		&\mbox{subject to: } V\uw=\tilde{\uzeta},\\
		&\left(
		\begin{array}{cc}
			-V & \uI_{M} \\
			V & \uI_{M} \\
		\end{array}
		\right)
		\left(
		\begin{array}{c}
			\uw \\
			\tilde{\uw} \\
		\end{array}
		\right)\geq
		\left(
		\begin{array}{c}
			-\tilde{\uzeta} \\
			\tilde{\uzeta} \\
		\end{array}
		\right),\\
		&~\mbox{and }\omega_j\geq0, j=1,\cdots,J_t,
	\end{split}
\end{align}	
where $M$ is the number of estimated moments, $V=(v_{ij})\in\mathbb{R}^{M\times J_t}$ with element $v_{ij}=t_j^i$, and $\tilde{\uzeta}=(\tilde{\zeta}_1,\cdots,\tilde{\zeta}_M)\trans\in\mathbb{R}^{M}$. Therefore, the largest eigenvalue is estimated by $\widehat{\lambda_{\max}(\Sigma_{\eta})}$ in (\ref{eigenmax_hat}) with $\hat{\uw}=(\hat{\uw}_1,\cdots,\hat{\uw}_{J_t})\trans$ minimizing LP (\ref{mpmom}).

\begin{theorem}\label{th:lr2}
	If conditions {\rm(\textbf{A.1})-(\textbf{A.8})} in the Supplementary Material hold and $p_1^{1-\nu}=o(n)$ for every positive constant $\nu$, $H$ is known, and the estimated largest eigenvalue $\widehat{\lambda_{\max}(\Sigma_{\eta})}$ is given by (\ref{eigenmax_hat}) with $\hat{\uw}$ minimizing the LP (\ref{mplp}) or (\ref{mpmom}), we have that under $H_0$ in (\ref{test}),
	\begin{align*}
		\lim_{n\rightarrow\infty}P\left(T_{n,p}>\frac{\sigma^2z_{\alpha}}{n}\sqrt{2\tr(\Sigma_{\eta}^2)}\right)
		\leq \alpha,
	\end{align*}
	and under $H_1$ in (\ref{test}),
	\begin{align*}
		\lim_{n\rightarrow\infty}P\left(T_{n,p}>\frac{\sigma^2z_{\alpha}}{n}\sqrt{2\tr(\Sigma_{\eta}^2)}\right)
		=\begin{cases}
			\alpha,& \hbox{if } n\frac{\ubeta\trans\Sigma_{\eta}^2\ubeta-\delta_0^2\lambda_{\max}^2(\Sigma_{\eta})}{\sigma^2\sqrt{2\tr(\Sigma_{\eta}^2)}}=o(1), \\
			1, & \hbox{if } ~n\frac{\ubeta\trans\Sigma_{\eta}^2\ubeta-\delta_0^2\lambda_{\max}^2(\Sigma_{\eta})}{\sigma^2\sqrt{2\tr(\Sigma_{\eta}^2)}}\rightarrow+\infty.
		\end{cases}
	\end{align*}
\end{theorem}
With a preset significant level $\alpha$ , a statistician rejects $H_0$ if p-value $<\alpha$, where the p-value is immediately calculated by $\mbox{p-value} = 1-\Phi(nT_{n,p}(2\widehat{\sigma^4\tr(\Sigma_{\eta}^2)})^{-1/2})$.

\begin{algorithm}[htb]
\caption{HDTRD}
\label{alg:A}
\begin{algorithmic}[1]
\STATE \textbf{Input:} Dataset $\{(y_i, \ux_i,\uz_i),i=1,\cdots,n\}$, significant level $\alpha$.
\STATE Step {\bf1}. Calculate $S_n=\frac{1}{n}\sum_{i=1}^n\hat{\ueta}_i\hat{\ueta}_i\trans$, where $\hat{\ueta}_i=\ux_i-\widehat{H}\uz_i$ with $\widehat{H}$ in (\ref{obj_phi}).		
\STATE Step {\bf2}. Calculate $\widehat{\lambda_{\max}(\Sigma_{\eta})}$ by (\ref{eigenmax_hat}), where $\hat{\uw}$ minimizes LP (\ref{mplp}) or (\ref{mpmom}).	
\STATE Step {\bf3}. Calculate the test statistic $T_{n,p}$ by (\ref{statistic}).
\STATE Step {\bf4}. Calculate $\mbox{p-value} = 1-\Phi(nT_{n,p}(2\widehat{\sigma^4\tr(\Sigma_{\eta}^2)})^{-1/2})$.
\STATE \textbf{Output:} p-value, or rejecting $H_0$ if p-value $<\alpha$.
\end{algorithmic}
\end{algorithm}

\subsection{Implementation}
In Section \ref{sec:eigmax}, we have proposed methods to estimate consistently the largest eigenvalue of $\Sigma_{\eta}$ based on $S_n=n^{-1}\sum_{i=1}^{n}\ueta_i\ueta_i\trans$ when $H$ is known. However, $H$ is unknown in practice, which results in that $S_n$ is not available. It is natural to replace $\ueta_i$ by its estimator $\hat{\ueta}_i=\ux_i-\widehat{H}\uz_i$, where $\widehat{H}$ is a consistent estimator of $H=\Sigma_{xz}\Sigma_z^{-1}$. Thus, we can replace $S_n$ by
$\widehat{S_n}=n^{-1}\sum_{i=1}^{n}\hat{\ueta}_i\hat{\ueta}_i\trans$, which is the sample covariance of $\{\hat{\ueta}_1,\cdots,\hat{\ueta}_n\}$.
In the high-dimensional setting, 
we can consistently estimate $H$ by minimizing the penalized quadratic loss
\begin{align}\label{obj_phi}
	\widehat{H}=\argmin_{H\in\mathbb{R}^{p_1\times p_2}}\frac{1}{2n}\sum_{i=1}^{n}\|\ux_i-H\uz_i\|^2+\sum_{j=1}^{p_1}\lambda_w\|\uh_j\|_1,
\end{align}
where $\uh_j$ is the $j$th column of $H\trans$.
We summarize the new test procedure in Algorithm \ref{alg:A}, named ``HDTRD". In Appendix G.1 of the Supplementary Material, we conduct a simulation study to investigate the performance of estimation procedures for $\lambda_{\max}(\Sigma_{\eta})$, from which it is seen that the proposed methods estimate $\lambda_{\max}(\Sigma_{\eta})$ well.

\section{Test of transferability}\label{sec:test_trans}
\subsection{Unified transfer learning}\label{sec:utr}
Let $\uy_k=(y_{k1},\cdots,y_{kn_k})\trans$, $\uX_k=(\ux_{k1},\cdots,\ux_{kn_k})\trans$ and $\ueps_k=(\eps_{k1},\cdots,\eps_{kn_k})\trans$ for $k=0,1,\cdots,K_s$, where $\{y_{ki},i=1,\cdots,n_k\}$ and $\{\eps_{ki},i=1,\cdots,n_k\}$ are the responses and errors from the $k$th dataset $\uV_k$, respectively. Here $\uV_0=\{(y_{0i},\ux_{0i}),i=1,\cdots,n_0\}$ denotes the target dataset, and $\uV_k=\{(y_{ki},\ux_{ki}),i=1,\cdots,n_k\}$ denotes $k$th source dataset for $k=1,\cdots,K_s$. We consider following linear regression model for the $k$th dataset,
\begin{align}\label{reg_transfer}
	y_{ki} = \ux_{ki}\trans\ubeta_k+\eps_{ki}, ~i=1,\cdots n_k.
\end{align}	

Let $N=\sum_{k=0}^{K_s}n_k$.
Define $\mathbb{Y}=(\uy_0\trans,\uy_1\trans,\cdots,\uy_{K_s}\trans)\trans\in\mathbb{R}^{N}$, $\mathbb{E}=(\ueps_0\trans,\ueps_1\trans,\cdots,\ueps_{K_s}\trans)\trans\in\mathbb{R}^{N}$, $\uB=(\ubeta_0\trans,(\ubeta_1-\ubeta_0)\trans,\cdots,(\ubeta_{K_s}-\ubeta_0)\trans)\trans\in\mathbb{R}^{p(K_s+1)}$, and
\begin{align*}
	\mathbb{X}
	=\left(
	\begin{array}{cccccc}
		\uX_0 & \uzero & \cdots & \uzero \\
		\uX_1 & \uX_1 &  &  \\
		\vdots &  & \ddots &  \\
		\uX_{K_s} &\uzero  &\cdots  & \uX_{K_s} \\
	\end{array}
	\right)\in\mathbb{R}^{N\times (pK_s+p)}.
\end{align*}
\cite{Liu2024} proposed a unified procedure for transfer learning in high-dimensional linear regression model
\begin{align}\label{model_lr}
	\mathbb{Y} = \mathbb{X}\uB + \mathbb{E}.
\end{align}
A penalized quadratic loss function is proposed by \cite{Liu2024}
\begin{align}\label{loss_utr}
	L(\uB)=\frac{1}{2N}\l\| \mathbb{Y} - \mathbb{X}\uB \r\|^2+\lambda\|\uB\|_1,
\end{align}
where $\lambda$ is a tuning parameter.
Then, the first $p$ elements of $\widehat{\uB}$ yield the estimator $\hubeta_0$ of interest, where $\widehat{\uB}$ is the minimizer of the loss function $L(\uB)$ in (\ref{loss_utr}).

\subsection{Transferability testing}\label{sec:test_trans_sub}

For $k=1,\cdots, K_s$, according to the difference of two linear regression models in (\ref{reg_transfer}) based on data $\uV_0$ and $\uV_k$, we construct a new linear regression model
\begin{align}\label{model_lr_0k}
	y_{0ki} = \ux_{ki}\trans(\ubeta_0-\ubeta_k)+\ux_{0ki}\trans\ubeta_0+\eps_{0ki}, ~i=1,\cdots, \check{n}_k,
\end{align}
where $\check{n}_k=\min\{n_0,n_k\}$, $y_{0ki}=y_{0i}-y_{ki}$, $\ux_{0ki}=\ux_{0i}-\ux_{ki}$, and $\eps_{0ki}=\eps_{0i}-\eps_{ki}$, $i=1,\cdots, \check{n}_k$.
To detect the transferability of the $k$th source dataset $\uV_k$, it is of interest to test traditionally whether the contrast coefficient $\ubeta_k-\ubeta_0$ is $\uzero$ or not.
From literature \citep{Tian2023,Chen2023}, researchers turn the hypothesis test of transferability into the testing problem
\begin{align}\label{test_classic_k}
	H_0^{c,k}: \ubeta_k-\ubeta_0=\uzero \quad \mbox{versus}\quad H_1^{c,k}: \ubeta_k-\ubeta_0 \neq \uzero.
\end{align}
In model (\ref{model_lr_0k}), the above testing problem is equivalent to the testing problem (\ref{test_classic}).

In practice, since $\ubeta_0$ is unknown, we can consistently estimate $\ubeta_0$ by minimizing the penalized quadratic loss based on the pooled data from $\uV_0$ and $\uV_k$,
\begin{align}\label{obj_beta0k_a}
	\hubeta_{0,k}^{ini}=\argmin_{\ubeta\in\mathbb{R}^{p}}\frac{1}{2\check{n}_k}\sum_{i=1}^{\check{n}_k}(y_{0ki}-\ux_{0ki}\trans\ubeta)^2+\lambda\|\ubeta\|_1,
\end{align}
where $\lambda$ is a tuning parameter.
Thus, based on $\hubeta_{0,k}^{ini}$, we propose a new test statistic for the classic testing problem (\ref{test_classic_k})
\begin{align}\label{Tnp_0k_a}
	T_{n,p}^{0,k}
	=\frac{1}{\check{n}_k(\check{n}_k-1)}
	\sum_{i\neq j}\ux_{ki}\trans\ux_{kj}(y_{0ki}-\ux_{0ki}\trans\hubeta_{0,k}^{ini})(y_{0kj}-\ux_{0kj}\trans\hubeta_{0,k}^{ini}).
\end{align}
For each $k=1,\cdots,K_s$, we define a family of contrast coefficients for the $k$th model
\begin{multline*}
	\mathfrak{F}_k=\bigg\{\ubeta\in\mathbb{R}^{p}:\ubeta\trans\Sigma_{\eta,k}\ubeta=o(1), \ubeta\trans\Sigma_{\eta,k}^2\ubeta=o\left(\frac{\tr(\Sigma_{x,k}^2)}{\varsigma_{k,n}}\right),\\
	\ubeta\trans\Sigma_{\eta,k}\Sigma_{x,k}\Sigma_{\eta,k}\ubeta=o\left(\frac{\tr(\Sigma_{x,k}^2)}{\check{n}_k}\right)\bigg\},
\end{multline*}
where $\varsigma_{k,n}=\check{n}_ks_{k,\phi}\log p$ with $s_{k,\phi}=\|\ubeta_{k,\phi}\|_0$ and $\ubeta_{k,\phi}$ defined in Lemma F1 in the Supplementary Material, and $\Sigma_{\eta,k}=\Sigma_{x,k}-\Sigma_{x,k}\left(\Sigma_{x,0}+\Sigma_{x,k}\right)^{-1}\Sigma_{x,k}$ is the covariance of $\ueta_{ki}=\ux_{ki}-H_k\ux_{0ki}$ with $H_k=\Sigma_{x,k}\left(\Sigma_{x,0}+\Sigma_{x,k}\right)^{-1}$.
The asymptotic normality of $T_{n,p}^{0,k}$ is established under the null and alternative hypotheses in following Theorem \ref{th:lr_0k_a}.
\begin{theorem}\label{th:lr_0k_a}
	If conditions {\rm(\textbf{B.1})-(\textbf{B.5})} in the Supplementary Material hold,
	it follows that for any given $\ubeta_k-\ubeta_0\in\mathfrak{F}_k$, $k=1,\cdots,K_s$,
	\begin{align*}
		\frac{T_{n,p}^{0,k}-(\ubeta_k-\ubeta_0)\trans\Sigma_{\eta,k}^2(\ubeta_k-\ubeta_0)}{\check{n}_k^{-1}\sigma_{0k}^2\sqrt{2\tr(\Sigma_{x,k}^2)}}\wcon \mathcal{N}(0,1),
	\end{align*}
	where $\sigma_{0k}^2=\sigma_0^2+\sigma_k^2$, and $\sigma_k^2$ is the variance of $\eps_{ki}$.
\end{theorem}

As demonstrated in Section \ref{sec:test_old}, the high correlation between $\ux_{0k}$ and $\ux_k$ break Condition (B.5). Thus, a projection method is required. As proposed in (\ref{hatH}), we estimate $H_k$ by
\begin{align}\label{hatH_a}
	\widehat{H}_k = \argmin_{H\in\mathbb{R}^{p\times p}}\frac{1}{2\check{n}_k}\sum_{i=1}^{\check{n}_k}\|\ux_{ki}-H\ux_{0ki}\|^2+\lambda\sum_{j=1}^{p}\|\uh_j\|_1.
\end{align}
Based on the residual of the projection score $\hat{\ueta}_{ki}=\ux_{ki}-\widehat{H}_k\ux_{0ki}$ for $i=1,\cdots,\check{n}_k$, we construct a test statistic $T_{n,p}^{0,k,Proj}$ as
\begin{align}\label{Tnp_project_a}
	T_{n,p}^{0,k,Proj}
	=\frac{1}{\check{n}_k(\check{n}_k-1)}
	\sum_{i\neq j}\hat{\ueta}_{ki}\trans\hat{\ueta}_{kj}(y_{0ki}-\ux_{0ki}\trans\hubeta_{0,k}^{ini})(y_{0kj}-\ux_{0kj}\trans\hubeta_{0,k}^{ini}).
\end{align}
For each $k=1,\cdots,K_s$, we define a family of contrast coefficients for the $k$th dataset as $\mathfrak{F}_k^{Proj}=\{\ubeta\in\mathfrak{F}_k:\ubeta\trans\Sigma_{\eta,k}^2\ubeta=o(\varsigma_n^{-1}\tr(\Sigma_{\eta,k}^2))\}$.
We can show that the test statistic $T_{n,p}^{0,k,Proj}$ is asymptotically normally distributed in following Theorem \ref{th:lr_0k_proj_a}.
\begin{theorem}\label{th:lr_0k_proj_a}
	If conditions {\rm(\textbf{B.1})-(\textbf{B.7})} in the Supplementary Material hold,
	it follows that for any $\ubeta_k-\ubeta_0\in\mathfrak{F}_k^{Proj}$, $k=1,\cdots,K_s$,
	\begin{align*}
		\frac{T_{n,p}^{0,k,Proj}-(\ubeta_k-\ubeta_0)\trans\Sigma_{\eta,k}^2(\ubeta_k-\ubeta_0)}{\check{n}_k^{-1}\sigma_{0k}^2\sqrt{2\tr(\Sigma_{\eta,k}^2)}}\wcon \mathcal{N}(0,1).
	\end{align*}
\end{theorem}

For an index set $\mathcal{A}\subset \{1,\cdots,K_s\}$, denote by $\delta_0=\max_{k\in\mathcal{A}}\|\ubeta_k-\ubeta_0\|$ the transfer level of $\mathcal{A}$.
Denote by $\mathcal{A}_{\delta_0}=\left\{k\in\{1,\cdots,K_s\}: \|\ubeta_k-\ubeta_0\|\leq \delta_0\right\}$ the index set of transferable sources with transfer level $\delta_0$.
As pointed out in Theorem 1 of \cite{Tian2023}, when $\delta_0\ll (sn_0^{-1}\log p)^{1/2}$ and $n_0\ll \min_{k\in\mathcal{A}_{\delta_0}}n_k$, their algorithm produces an estimator with better converge rate than that in ordinary Lasso without using source dataset. This suggests that, instead of hypothesis test (\ref{test_classic_k}), it is better to turn the test of transferability into test of relevant difference,
\begin{align}\label{test_k}
	H_0^k: \|\ubeta_k-\ubeta_0\|\leq\delta_0 \quad \mbox{versus}\quad H_1^k: \|\ubeta_k-\ubeta_0\|>\delta_0,
\end{align}
which is equivalent to testing relevant difference (\ref{test}) in the linear regression model (\ref{model_lr_0k}).
We propose a new test statistic for testing problem (\ref{test_k})
\begin{align}\label{statistic_k}
	T_{n,p}^{k,Proj} = T_{n,p}^{0,k,Proj}-\delta_0^2\widehat{\lambda_{\max}^2(\Sigma_{\eta,k})},
\end{align}
where $\widehat{\lambda_{\max}(\Sigma_{\eta,k})}$ is the estimator of the largest eigenvalue of $\Sigma_{\eta,k}$ based on
$\{\hat{\ueta}_{ki},i=1,\cdots,\check{n}_k\}$. Based on test statistic (\ref{statistic_k}), we summarize the new transfer learning procedure in Algorithm \ref{alg:B} (``TUTrans").
When $H_k$ is known, $\{\ueta_{ki},i=1,\cdots,\check{n}_k\}$ is i.i.d.. We establish the asymptotic properties for $T_{n,p}^{k,Proj}$ with known $H_k$ in Theorem \ref{th:lr_k2} for $k=1,\cdots,K_s$.
\begin{theorem}\label{th:lr_k2}
	If conditions {\rm(\textbf{B.1})-(\textbf{B.8})} in the Supplementary Material hold and $p^{1-\nu}=o(n_k)$ for every positive constant $\nu$, $H_k$ is known, and the estimated largest eigenvalue $\widehat{\lambda_{\max}(\Sigma_{\eta,k})}$ is given by (\ref{eigenmax_hat}) with $\hat{\uw}$ minimizing the linear programming problem (\ref{mplp}) or (\ref{mpmom}), we have that under $H_0^k$ in (\ref{test_k}),
	\begin{align*}		\lim_{\check{n}_k\rightarrow\infty}P\left(T_{n,p}^{k,Proj}>\frac{\sigma_{0k}^2z_{\alpha}}{\check{n}_k}\sqrt{2\tr(\Sigma_{\eta,k}^2)}\right)
		\leq \alpha,
	\end{align*}
	and under $H_1^k$ in (\ref{test_k}),
	\begin{align*}		
		\lim_{\check{n}_k\rightarrow\infty}&P\left(T_{n,p}^{k,Proj}>\frac{\sigma_{0k}^2z_{\alpha}}{\check{n}_k}\sqrt{2\tr(\Sigma_{\eta,k}^2)}\right)\\
		&=\begin{cases}
			\alpha,& \hbox{if } \check{n}_k\frac{(\ubeta_k-\ubeta_0)\trans\Sigma_{\eta,k}^2(\ubeta_k-\ubeta_0)-\delta_0^2\lambda_{\max}^2(\Sigma_{\eta,k})}{\sigma_{0k}^2\sqrt{2\tr(\Sigma_{\eta,k}^2)}}=o(1), \\
			1, & \hbox{if } ~\check{n}_k\frac{(\ubeta_k-\ubeta_0)\trans\Sigma_{\eta,k}^2(\ubeta_k-\ubeta_0)-\delta_0^2\lambda_{\max}^2(\Sigma_{\eta,k})}{\sigma_{0k}^2\sqrt{2\tr(\Sigma_{\eta,k}^2)}}\rightarrow+\infty.
		\end{cases}
	\end{align*}
\end{theorem}

\cite{Liu2024} considered a linear regression model for the contrast coefficient $\ubeta_k-\ubeta_0$
\begin{align}\label{model_lr_k}
	\mathbb{Y}_k = \mathbb{X}_k(\ubeta_k-\ubeta_0) + \mathbb{Z}_k\ubeta_0 + \mathbb{E}_k,\quad k=1,\cdots,K_s,
\end{align}
where $\mathbb{Y}_k=(\uy_0\trans,\uy_k\trans)\trans\in\mathbb{R}^{n_0+n_k}$, $\mathbb{X}_k=(\uzero\trans, \uX_k\trans)\trans\in\mathbb{R}^{(n_0+n_k)\times p}$,  $\mathbb{Z}_k=(\uX_0\trans, \uX_k\trans)\trans\in\mathbb{R}^{(n_0+n_k)\times p}$, and $\mathbb{E}_k=(\ueps_0\trans,\ueps_k\trans)\trans\in\mathbb{R}^{n_0+n_k}$.
Based on model (\ref{model_lr_k}), \cite{Liu2024} borrowed the test statistic $T_{n,p}^{0}$ in (\ref{Tnp}) to detect transferability for the source dataset $\uV_k$.
\cite{Liu2024} estimated $\ubeta_0$ by minimizing the penalized quadratic loss based on pooled data from $\uV_0$ and $\uV_k$,
\begin{align}\label{obj_beta0k}
	\tilde{\ubeta}_{0,k}^{ini}=\argmin_{\ubeta\in\mathbb{R}^{p}}\frac{1}{2(n_0+n_k)}\sum_{t\in\{0,k\}}\sum_{i=1}^{n_t}(y_{ti}-\ux_{ti}\trans\ubeta)^2+\lambda\|\ubeta\|_1.
\end{align}
\cite{Liu2024} proposed a test statistic for the classic testing problem (\ref{test_classic_k})
\begin{align}\label{Tnp_0k}
	\tilde{T}_{n,p}^{0,k}
	=\frac{1}{n_k(n_k-1)}\sum_{i\neq j}\ux_{ki}\trans\ux_{kj}(y_{ki}-\ux_{ki}\trans\tilde{\ubeta}_{0,k}^{ini})(y_{kj}-\ux_{kj}\trans\tilde{\ubeta}_{0,k}^{ini}).
\end{align}

In Appendix G.2 of the Supplementary Material, we conduct a simulation study to compare the performance of test statistics $T_{n,p}^{0,k,Proj}$ given in (\ref{Tnp_project_a}), $T_{n,p}^{0,k}$ given in (\ref{Tnp_0k_a}), and $\tilde{T}_{n,p}^{0,k}$ given in (\ref{Tnp_0k}). As expected, it is seen that $T_{n,p}^{0,k,Proj}$ controls Type-I error well, while both $T_{n,p}^{0,k}$ and $\tilde{T}_{n,p}^{0,k}$ fail to control Type-I error. $\tilde{T}_{n,p}^{0,k}$ has systematically higher Type-I error in all scenarios compared with $T_{n,p}^{0,k,Proj}$ and $T_{n,p}^{0,k}$, which is caused by linear correlation between $\mathbb{X}_k$ and $ \mathbb{Z}_k$ in model (\ref{model_lr_k}).
In Appendix G.3 of the Supplementary Material, we provide the cross validation method, as well as Algorithm \ref{alg:C}, to choose the transferable level $\delta_0$ in practice.

\begin{algorithm}[hbt!]
	\caption{TUTrans: UTrans incorporating the HDTRD.}
	\label{alg:B}
	\begin{algorithmic}[1]
		\STATE \textbf{Input:} Datasets $\{(\mathbb{Y}_k, \mathbb{X}_k),0\leq k\leq K_s\}$, transfer level $\delta_0$, significant level $\alpha$.
		\STATE Step {\bf1}. Obtain the estimator $\hubeta_{0,k}^{ini}$ by (\ref{obj_beta0k_a}) on data $\{\uV_0,\uV_k\}$ for $k=1,\cdots,K_s$.
		\STATE Step {\bf2}. Detect transferability of source data.
		\FOR{$k=1,2,\cdots,K_s$}
		\STATE Step {\bf2.1}. Construct the linear regression model (\ref{model_lr_0k}) replaced $\ubeta_0$ by $\hubeta_{0,k}^{ini}$.
		\STATE Step {\bf2.2}. Calculate $\hat{H}_k$ by (\ref{hatH_a}).
		\STATE Step {\bf2.3}. Calculate the test statistic $T_{n,p}^{k,Proj}$ given by (\ref{statistic_k}).
		\STATE Step {\bf2.4}. Make decision that $\uV_k$ is transferable and $k\in\hat{\mathcal{A}}_{\delta_0}$ if p-value $>\alpha$, where p-value is given by Algorithm \ref{alg:A} based on model (\ref{model_lr_0k}) on data $\uV_k$.
		\ENDFOR
		\STATE Step {\bf3}. Construct the linear regression model (\ref{model_lr}) based on $\{\uV_k,k\in\{0\}\cup\hat{\mathcal{A}}_{\delta_0}\}$.
		\STATE Step {\bf4}. Estimate $\hubeta_0$, the first $p$ elements of $\widehat{\uB}$ that minimizes $L(\uB)$ in (\ref{loss_utr}).
		\STATE \textbf{Output:} Transferable set $\hat{\mathcal{A}}_{\delta_0}$ and estimator $\hubeta_0$.
	\end{algorithmic}
\end{algorithm}

\section{Numerical studies}\label{sec:sim_test}
\subsection{Performance of test with high-dimensional control factors}\label{sec:sim_test_nuisan}
In this section, we consider the testing problem (\ref{test}) based on the linear regression model (\ref{reg.1}).
We generate covariate $\uu=(\ux\trans,\uz\trans)\trans$ from the multinormal distribution with covariance satisfying autoregressive structure that the sample $\uu_i$ are drawn from the $p$-variate normal distribution with mean zero and covariance matrix $
\Sigma = \{\sigma_{ij} = \rho^{|i-j|}, \ i,j = 1, \cdots, p\}$, where $\rho=0.6$ controls the strength of the correlation.

We consider the alternative hypothesis such that the coefficient structures have two different sparsity patterns: point sparsity and proportional sparsity. The former is the extremely sparse pattern with only a few nonzero coefficients, and the latter is the moderate sparse pattern with a certain proportion of nonzero coefficients. Suppose that the nonzero index set is $\calS = \{1, 2, \cdots, s\}, \ s < p$, and the magnitude of each coefficient is $\beta_j = \|\ubeta_{\calS}\|/\sqrt{s}$.
We take $s = 5$ for the point sparsity pattern, and $s = p_1/2$ for the proportional sparsity pattern. We set $p_1=p_2=p/2$, $\gamma_1=3$, $\gamma_2=1.5$, $\gamma_5=2$, and $\gamma_j=0$ for all $j\neq 1,2,5$.
We study the empirical power through changing the $\ell_2$-norm values of $\ubeta_{\calS}$. We study the performance of test procedure with high-dimensional control factors, and provide the performance of test procedure without control factors in Appendix G.4 of the Supplementary Material.

For comparison, we consider five methods: (1) the proposed test procedure with true $\lambda_{\max}(\Sigma_x)$, denoted by TRUE; (2) with $\widehat{\lambda_{\max}^{mpmo}(\Sigma_x)}$ by solving LP (\ref{mplp}), denoted by MAXM; (3) with $\widehat{\lambda_{\max}^{mplp}(\Sigma_x)}$ by solving LP (\ref{mpmom}), denoted by MAXL; (4) high-dimensional test proposed by \cite{Chen2023}, denoted by CLC; (5) high-dimensional test proposed by \cite{Yang2023}, denoted by YGZ. Here, $\widehat{\lambda_{\max}^{mpmo}(\Sigma_x)}$ and $\widehat{\lambda_{\max}^{mplp}(\Sigma_x)}$ denote the estimator of $\lambda_{\max}(\Sigma_x)$ by MPMO and MPLP with first four moments as constraints given in Section \ref{sec:eigmax}, respectively. To examine the performance in different scenarios, we consider sample size $n=200$, $p=300,700,1500$, and $\delta_0=c_0\sqrt{\log p/n}$ with $c_0=0, 0.5$ for the spare alternative and $c_0=0, 0.2$ for the dense alternative.

\begin{figure}[htb]
\centering
\begin{tabular}{cc}
\subfigure[Type-I error (sparse alternative)]{
\centering
\begin{minipage}[b]{0.48\linewidth}
\includegraphics[scale=0.30]{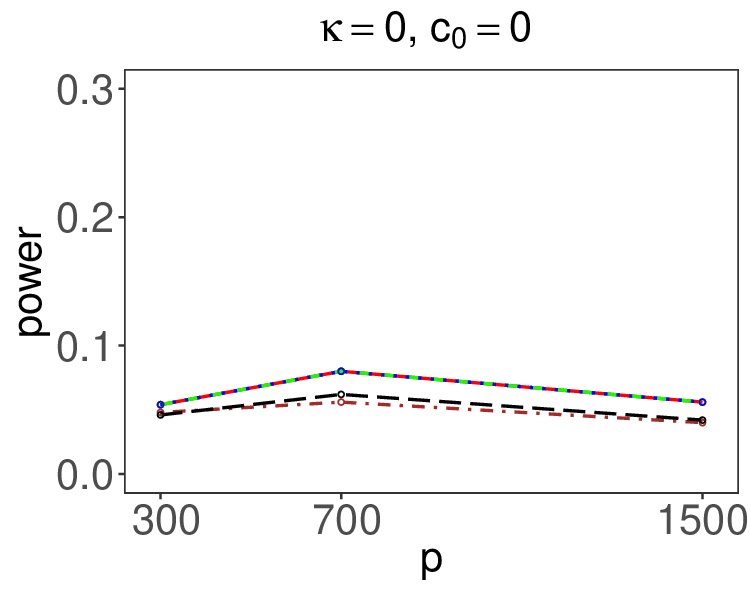}
\includegraphics[scale=0.30]{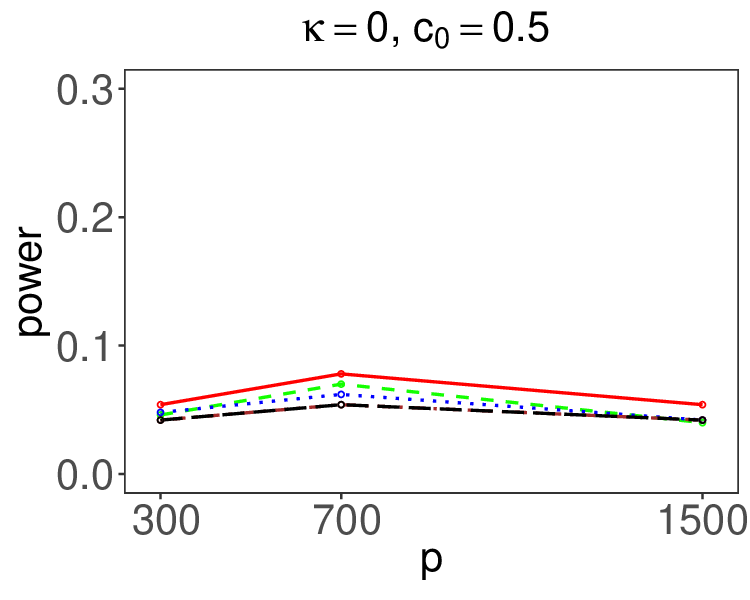}
\includegraphics[scale=0.30]{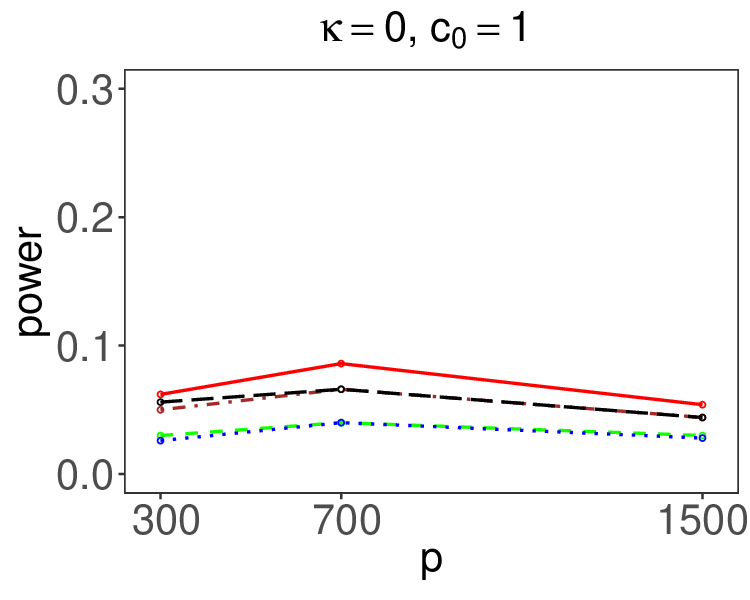}
\includegraphics[scale=0.30]{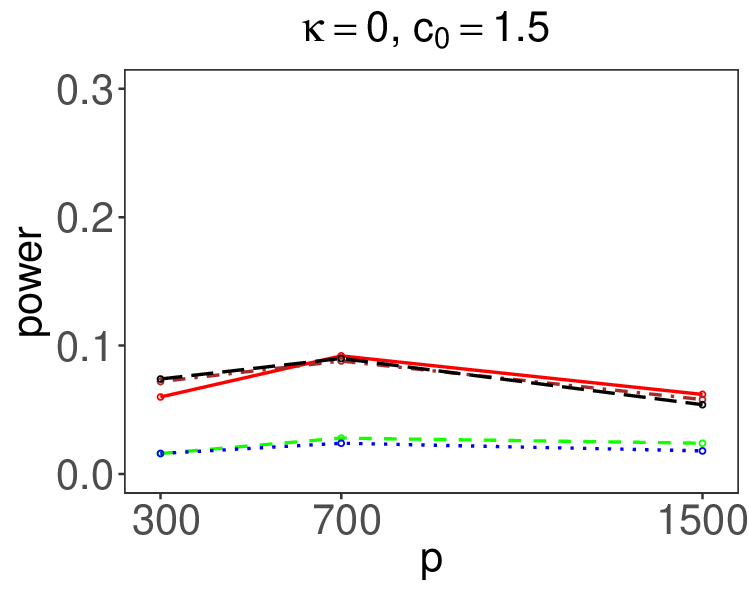}
\label{fig:type1_sparse1}
\end{minipage}}
&
\subfigure[Type-I error (dense alternative)]{
\centering
\begin{minipage}[b]{0.48\linewidth}
\includegraphics[scale=0.30]{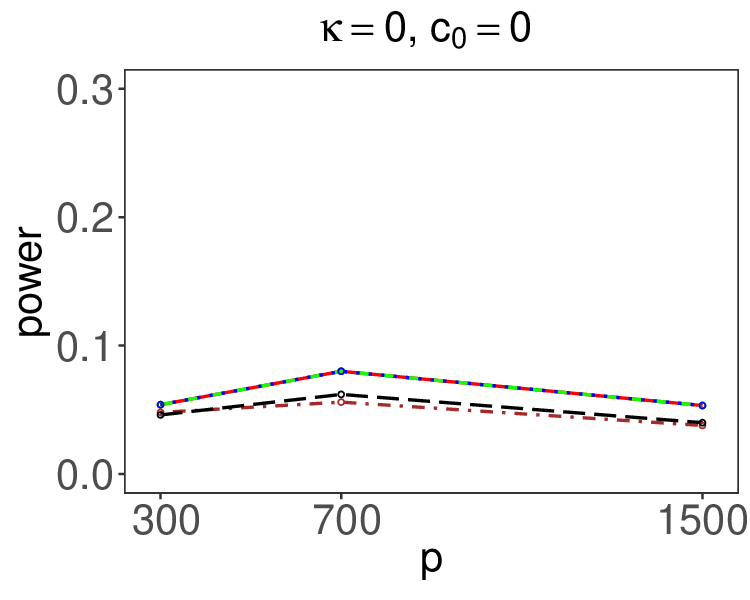}
\includegraphics[scale=0.30]{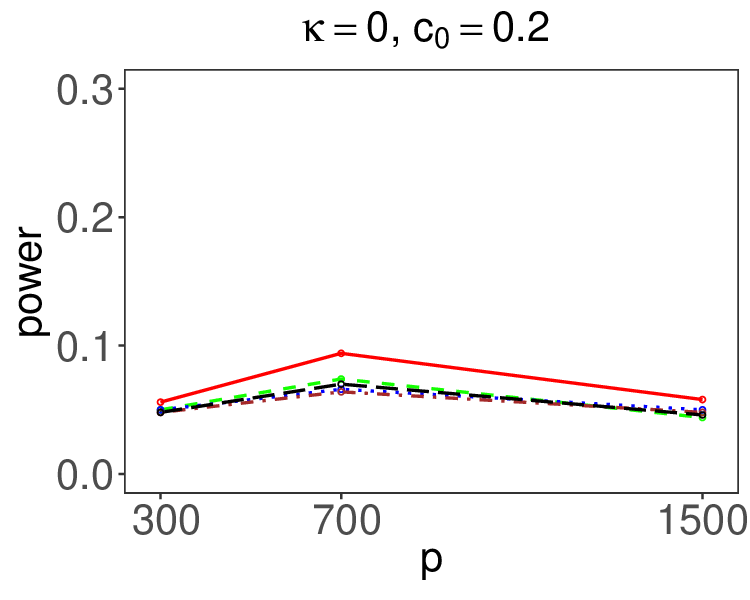}
\includegraphics[scale=0.30]{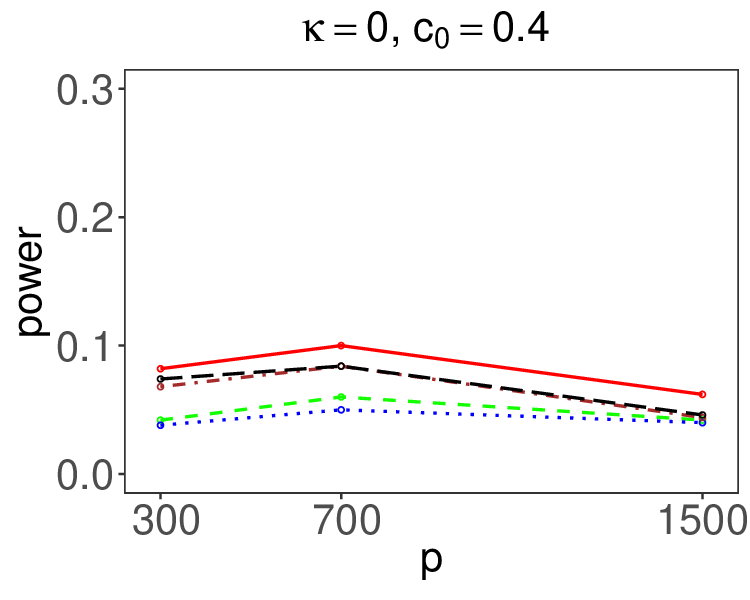}
\includegraphics[scale=0.30]{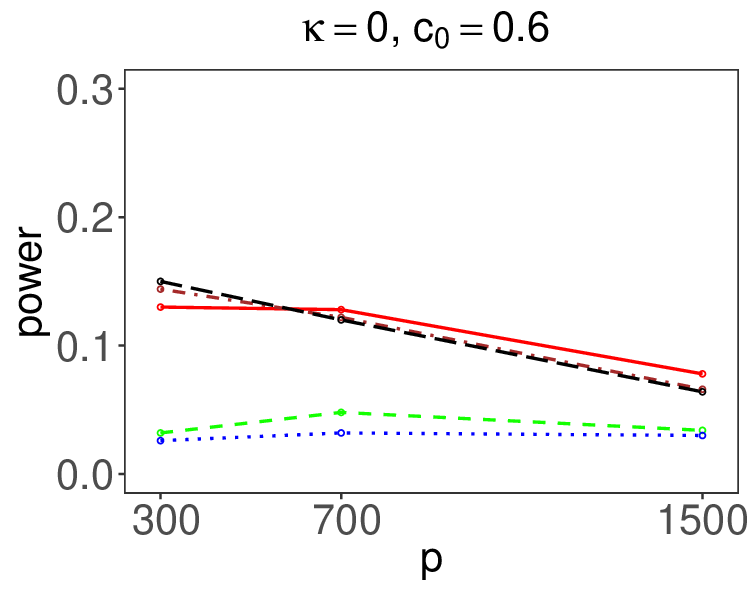}
\label{fig:type1_dense1}
\end{minipage}}
\\
\subfigure[Power (sparse alternative)]{
\centering
\begin{minipage}[b]{0.48\linewidth}                                                                            	
\includegraphics[scale=0.30]{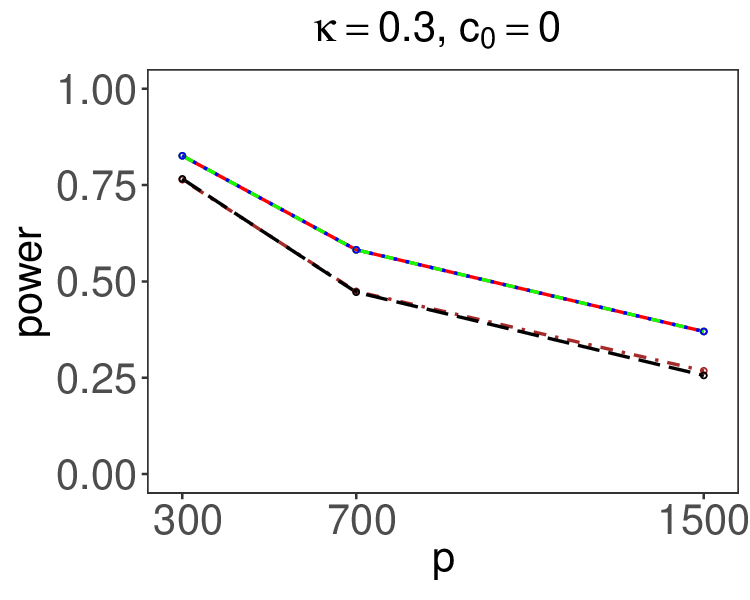}
\includegraphics[scale=0.30]{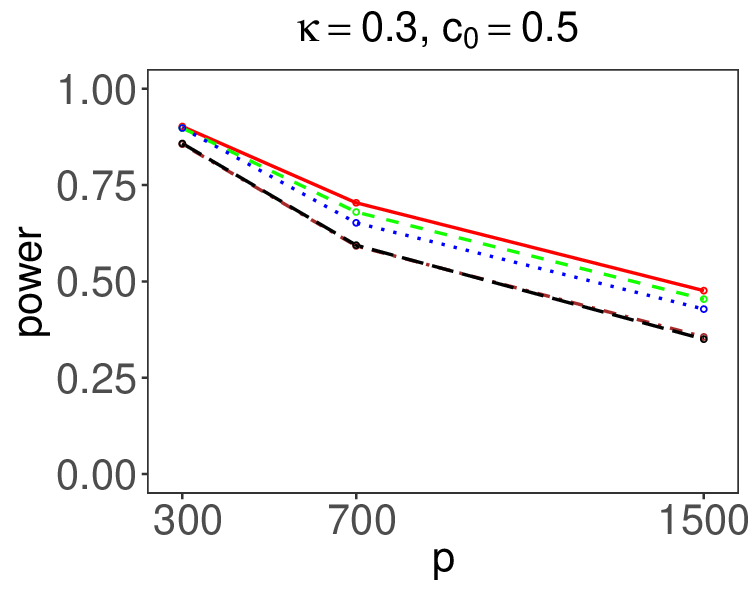}
\includegraphics[scale=0.30]{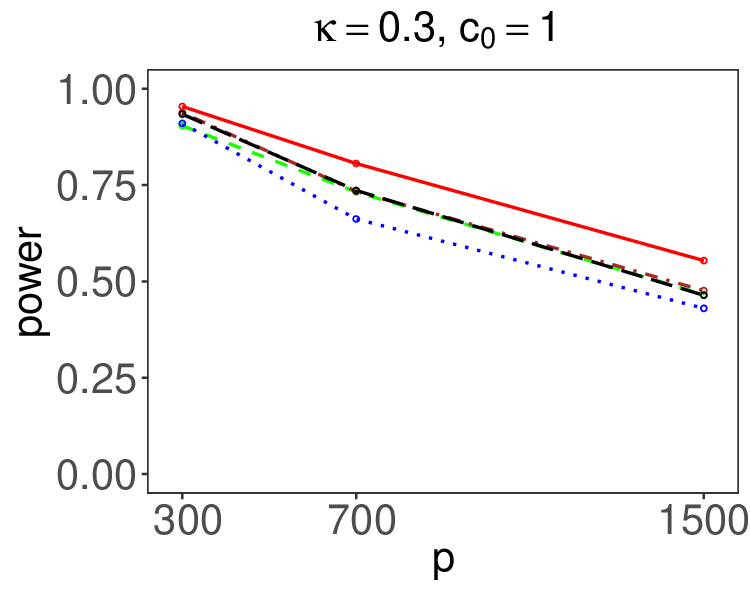}
\includegraphics[scale=0.30]{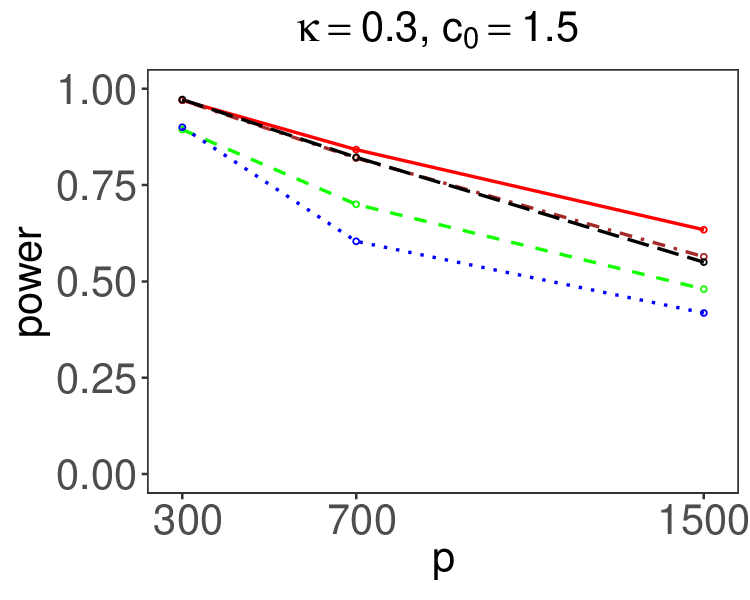}
\label{fig:power_sparse1}
\end{minipage}}
&
\subfigure[Power (dense alternative)]{
\centering
\begin{minipage}[b]{0.48\linewidth}
\includegraphics[scale=0.30]{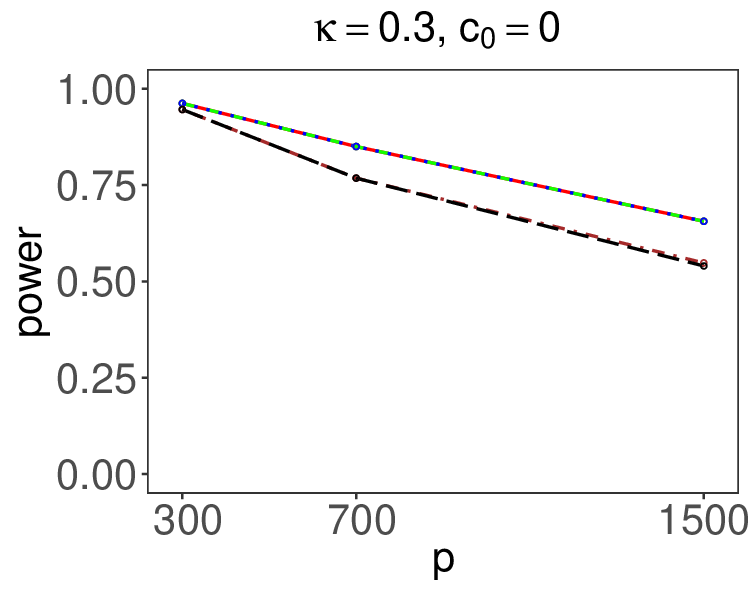}
\includegraphics[scale=0.30]{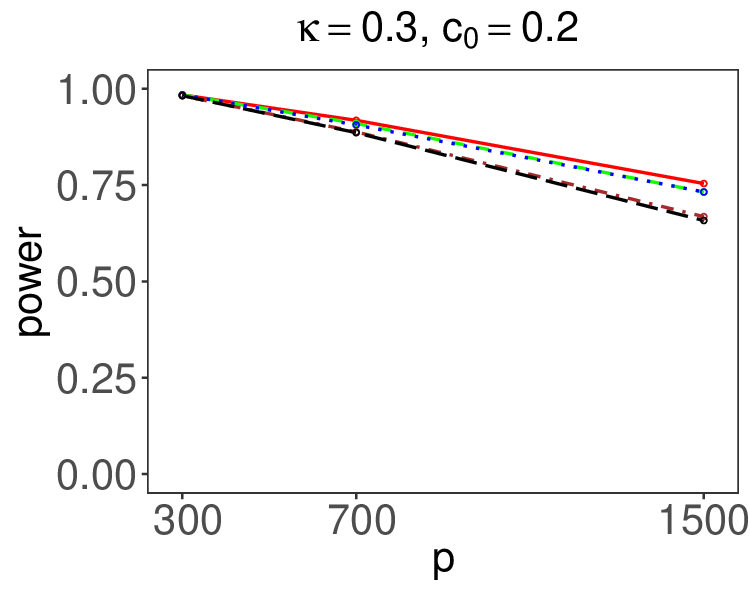}
\includegraphics[scale=0.30]{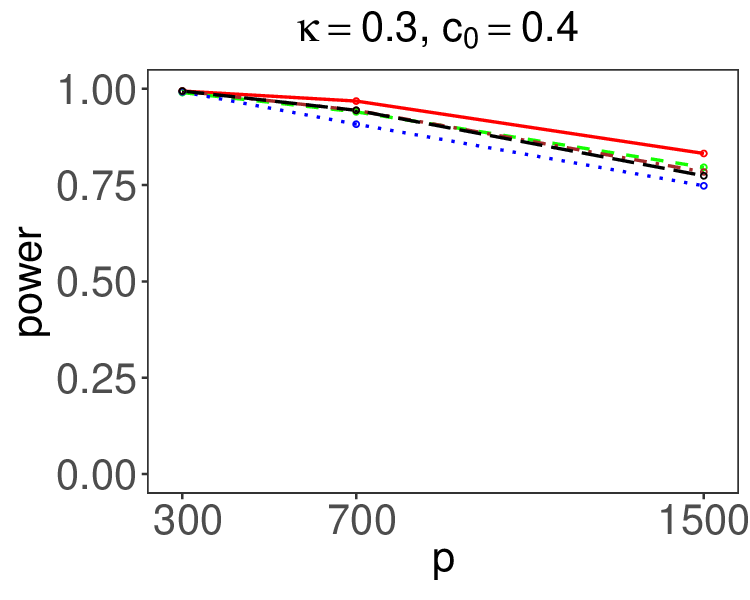}
\includegraphics[scale=0.30]{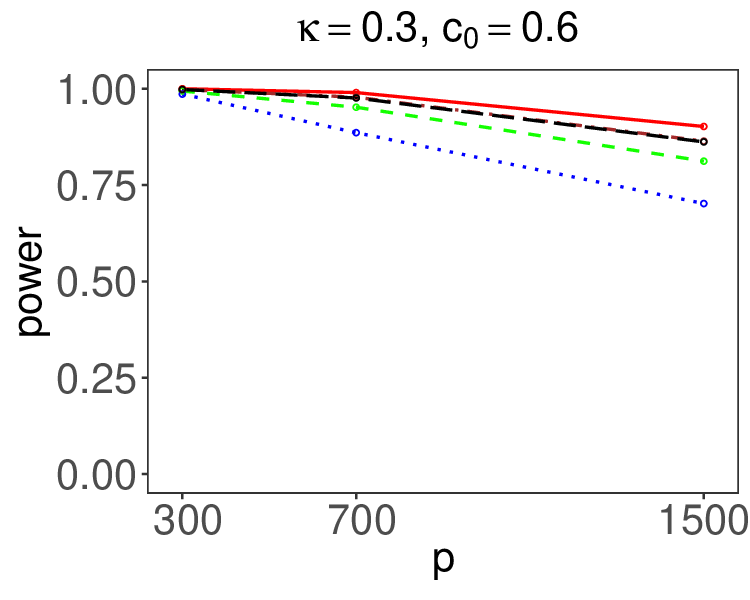}
\label{fig:power_dense1}
\end{minipage}}
\end{tabular}
\caption{Type-I errors and powers with control factors for 5 methods, including TRUE ({\color{red}red solid line}), MAXM ({\color{green}green dashed line}), MAXL ({\color{blue}blue dotted line}), CLC ({\color{brown}brown dotdash line}), and YGZ (black longdash line).}
\label{fig:control1}
\end{figure}

Figure \ref{fig:control1} depicts the Type-I error and power functions varying with $\|\ubeta\|=\kappa+\delta_0$, where $\kappa=0$ for the null hypothesis and  $\kappa=0.3$ for the alternative hypothesis. As expected, it is seen from Figure \ref{fig:control1} that the proposed methods can control Type-I error well, while Type-I errors of both YGZ and CLC become large and fail to control as $\delta_0$ grows up. All methods' powers decrease when dimension $p$ grows up, and increase when $\delta_0$ grows up.
Powers of the proposed methods are all lower than the CLC, but the difference is small when dimension $p$ is less than 1000.

\subsection{Performance of transfer learning}\label{sec:sim_hdtrans}
For the limit of space, we render the details of performance of transfer learning in Appendix G.5 of the Supplementary Material.	
It is seen from Figure G.3 that the proposed transfer learning methods incorporating two estimators of $\lambda_{\max}(\Sigma_x)$ achieve uniformally lower estimation error and prediction error than both UTR and GLMT.

\section{Genotype-tissue expression data}\label{sec:case_studies}
Alzheimer's disease (AD) 
is a brain disorder, and it gets worse over time. AD that causes the brain to shrink and brain cells to eventually die is characterized by changes in the brain that lead to deposits of certain proteins. AD is the most common cause of dementia, which is a gradual decline in memory, thinking, behavior and social skills. These changes affect badly a patient's ability to function. Genes play a crucial role in AD \citep{Gatz2006,Ricciarelli2017,Sims2020,Narasimhan2024,Blalock2004}, where late-onset AD shows heritability of 58-79\% and early-onset AD shows over 90\%. It is well-known that genetic association is a powerful and robust platform to build our understanding of the etiology of this complex disease. Apolipoprotein E (APOE) on chromosome 19 was the first risk gene identified as associating with late-onset AD \citep{Saunders1993}. APOE influences familial and early forms of disease, and it remains the strongest genetic risk factor. Study of APOE is motivated by its critical role in a myriad of human diseases. Genetic studies have revealed other three genes that may be linked to autosomal dominant or familial early onset AD. These three genes include amyloid precursor protein (APP), presenilin 1 (PSEN1), and presenilin 2 (PSEN2).


In this section, we consider the Genotype-Tissue Expression (GTEx) data\footnote{available at \href{https://gtexportal.org/}{https://gtexportal.org/}}, where GTEx measures gene expression levels from 48 tissues of 620 human donors. We focus on genes down-regulated in brain from patients with AD in a human gene set\footnote{named as {\it \small KEGG\_ALZHEIMERS\_DISEASE}} in the Human Molecular Signatures Database\footnote{available at \href{https://www.gsea-msigdb.org/gsea/msigdb/human/collections.jsp}{https://www.gsea-msigdb.org/gsea/msigdb/human/collections.jsp}}, which is related to the pathogenesis of incipient AD \citep{Blalock2004}. There are 166 genes in this gene set. After removing genes with missing values, there are 13 brain tissues with 119 genes. The details with tissue names and sample size of each tissue are provided in Appendix H in the Supplementary Material. We consider brain tissue {\it Hippocampus} as target data ($n_0=111$) in each experiment, and the rest 12 tissues as source datasets with $K_s=12$.
We study the association between APOE (or APP, PSEN1, PSEN2) as response and the rest genes in this gene set  as predictors.

For seeking stability, in each experiment we randomly split the samples of the target data into 20\% and 80\% with 100 times, where 80\% samples are used to estimate the coefficient $\ubeta_0$ in the target model, and the rest 20\% samples are used to calculate the prediction errors. The five-fold cross-validation method is used to select the transferable level $\delta_0$ from candidate $c_0\in\{1,\cdots,20\}$. We construct the test statistic and fit the model with four folds of sample, and validate with the rest of samples.

Table \ref{pe_apoe} lists the averages of prediction errors ($\mbox{PE}=\frac{1}{n_{new}}\sum_{i=1}^{n_{new}}(y_{new,i}-\uX_{new,i}\trans\hubeta_0)^2$) and number of identified source datasets over 100 random splits. For comparison, we consider 11 methods, including (1) UTrans (UTR, \cite{Liu2024}); (2) UTR with MAXM to select transferable source dataset (MAXM$_{\mbox{utr}}$); (3) UTR with MAXL to select transferable source dataset (MAXL$_{\mbox{utr}}$); (4) Trans-GLM (GLMT, \cite{Tian2023}); (5) GLMT with MAXM to select transferable source dataset (MAXM$_{\mbox{glmt}}$); (6) GLMT with MAXL to select transferable source dataset (MAXL$_{\mbox{glmt}}$); (7) Trans-Lasso with MAXM to select transferable source dataset (MAXM$_{\mbox{glmt}}$); (8) Trans-Lasso with MAXL to select transferable source dataset (MAXL$_{\mbox{glmt}}$); (9) Trans-Lasso (TLasso, \cite{Li2022Minimax}); (10) Lasso only on target dataset (NLasso); (11) Lasso on pooled dataset (TLasso). From Table \ref{pe_apoe}, we can see that the proposed methods MAXM and MAXL achieve uniformly smaller prediction error and loss compared with competitors UTR, GLMT, TLasso, NLasso and PLasso, respectively. The proposed methods MAXM and MAXL detect less source datasets than the GLMT and TLasso, which implies that the GLMT and TLasso yield the larger prediction error due to including negative-transfer source dataset. The UTR select comparable
number of source datasets with the proposed methods. As demonstrated in Section \ref{sec:test_trans_sub}, since the UTR fails to control type-I error, negative transfer causes higher prediction error.

\begin{table}
	\caption{Averages of prediction errors (PE) and number of identified source datasets ($\hat{K}_s$) over 100 random splits. 
	}
	\resizebox{\textwidth}{!}{
		\begin{tabular}{rrcccccccccccc}
			\hline
			\multirow{2}{*}{Gene}& &\multicolumn{11}{c}{Prediction error}\\
			\cline{3-13}
			&&MAXM$_{\mbox{utr}}$ & MAXL$_{\mbox{utr}}$ & UTR &MAXM$_{\mbox{glmt}}$ &MAXL$_{\mbox{glmt}}$ & GLMT & MAXM$_{\mbox{tLasso}}$ & MAXL$_{\mbox{tLasso}}$ & TLasso & NLasso &PLasso \\
			\cline{3-13}
			\hline
			APOE          &    PE&            0.556 & 0.563 & 0.567 & 0.518 & 0.521 & 0.542 & 0.688 & 0.695 & 0.700 & 0.566 & 0.757 \\
			&    $\hat{K}_s$&   6.470 & 6.520 & 9.940 & 6.470 & 6.520 & 11.100 & 6.470 & 6.520 & 12 & 0 & 12 \\
			[2ex]
			APP           &    PE&            0.460 & 0.458 & 0.473 & 0.536 & 0.535 & 0.548 & 0.536 & 0.522 & 0.513 & 0.545 & 0.698 \\
			&    $\hat{K}_s$&   6.470 & 6.430 & 8.140 & 6.470 & 6.430 & 11.180 & 6.470 & 6.430 & 12 & 0 & 12 \\
			[2ex]
			PSEN1         &    PE&            0.928 & 0.926 & 0.947 & 0.921 & 0.920 & 0.929 & 0.872 & 0.853 & 0.927 & 0.991 & 0.876 \\
			&    $\hat{K}_s$&   7.300 & 7.310 & 6.240 & 7.300 & 7.310 & 11.780 & 7.300 & 7.310 & 12 & 0 & 12 \\
			[2ex]
			PSEN2         &    PE&            0.834 & 0.844 & 0.857 & 1.019 & 1.020 & 1.029 & 0.998 & 0.999 & 1.019 & 1.029 & 1.017 \\
			&    $\hat{K}_s$&   9.690 & 9.740 & 7.670 & 9.690 & 9.740 & 11.880 & 9.690 & 9.740 & 12 & 0 & 12 \\
			\hline
		\end{tabular}
	}
	\label{pe_apoe}
\end{table}

\section{Conclusion}\label{sec:conclusion}
This paper studies hypothesis testing for the relevant difference of coefficients in high-dimensional linear regression models. This testing problem with the null hypothesis of no relevant difference between $\ubeta$ and $\uzero$ is $H_0: \|\ubeta\|\leq \delta_0$, instead of the classical hypothesis testing problem $H_0^{c}: \ubeta=\uzero$.  We propose a novel test procedure by overcoming the challenge of the estimation of the largest eigenvalue of a high-dimensional covariance matrix with assistance of the random matrix theory. It is natural to apply the test of relevant difference of coefficients to detect the transferability of source data in the transfer learning framework. We develop a testing algorithm, and furthermore, a transfer learning algorithm for estimation and prediction that is adaptive to the unknown informative set.

The classical hypothesis testing problem $H_0^{c}: \ubeta=\uzero \mbox{ versus } H_1^{c}: \ubeta \neq \uzero$ has been investigated in the high-dimensional generalized linear models (GLMs) and high-dimensional quantile regression models.
In the further work, using our similarity characterization, the test procedure of relevant difference of coefficients in other nonlinear regression models can be formulated.
Its application to detect the transferability of source data in the transfer learning beyond linear models remains an open problem, and is worthy of investigation under different characterizations for the similarity testing procedure.

\section*{Acknowledgements}
Liu's research is supported by the National Natural Science Foundation of China (12271329, 72331005).
%
%


\begin{center}
	{\large\bf SUPPLEMENTARY MATERIAL}
\end{center}
Supplementary Material includes the proof of Theorems and Lemmas, additional information for simulation studies and case study.


\end{document}